\pdfoutput=1
\documentclass[twocolumn,showpacs,preprintnumbers,amsmath,amssymb,floatfix]{revtex4}

\usepackage{graphicx}
\usepackage{amsmath,amssymb,combelow} 
\usepackage{epsfig}
\usepackage{dcolumn}
\usepackage{bm}
\usepackage{rotating}
\usepackage{times}
\usepackage{units}
\usepackage{hyperref}
\begin{document}


\title{Superconductivity in Hydrogenated Graphites}
\author{Nadina Gheorghiu}
\email{Nadina.Gheorghiu@yahoo.com}
\affiliation{UES Inc., Dayton, OH 45432}  
\author{Charles R. Ebbing}
\affiliation{The University of Dayton Research Institute, Dayton, OH 45469}
\author{Timothy J. Haugan}
\affiliation{The Air Force Research Laboratory (AFRL), Aerospace Systems Directorate, AFRL/RQ, Wright-Patterson AFB, OH 45433}
            
\date{\today}

\begin{abstract}
{We report transport and magnetization measurements on graphites that have been hydrogenated by intercalation with an alkane (octane).
The temperature-dependent electrical resistivity shows anomalies manifested as reentrant insulator-metal transitions. 
Below $T \sim 50$ K, the magnetoresistance data shows both antiferromagnetic (AFM) and ferromagnetic (FM) behavior as the 
magnetic field is decrease or increased, respectively. The system is possibly an unconventional magnetic superconductor.
The irreversibility observed in the field-cooled vs. the zero-field cooled data 
for a sufficiently high magnetic field suggests that the system might enter a superconducting state below $T_{c} \sim 50$ K. 
Energy gap data is obtained from nonlocal electric differential conductance measurements. 
An excitonic mechanism is likely driving the system to the superconducting state below the same $T \sim 50$ K, where the gap is divergent.
We find that the hydrogenated carbon fiber is a multiple gap system with \textit{critical temperatures estimates above room temperature}. 
The temperature dependence of the superconducting gap follows the flat-band energy relationship, with  
the flat band gap parameter linearly increasing with the temperature above $T_{c} \sim 50$ K.
Thus, we find that either a magnetic or an electric field can drive this hydrogenated graphitic system to superconducting state below $T_{c} \sim 50$ K.
In addition, AF spin fluctuations creates pseudogap states above $T_{c} \sim 50$ K.}
\end{abstract}

\maketitle 

\section{Introduction}
The liquefaction of helium was a major technological achievement that further allowed Kamerlingh Onnes to observe superconductivity (SC) as the
nondissipative flow of electricity through mercury below the critical temperature $T_{c} = 4.19$ K \cite{Onnes}.
Likewise, fundamental discoveries in atomic physics that lead to the laws of quantum mechanics allowed 
Bardeen, Cooper, and Schrieffer to create a microscopic description of the new phenomenon known as the BCS theory \cite{BCS}.
SC still is among the most daunting research topics in condensed matter physics, 
as the discovery of YBCO certainly showed \cite{Ashburn}.
New SC materials, including high-$T_{c}$ SC (HTS) are being found and classified \cite{Chu1,Buzea}.
Carbon(C)-based SC materials \cite{Haruyama} can be graphite intercalated compounds \cite{Dresselhaus}.
More recently, room-temperature SC in graphitic systems like highly oriented pyrolytic graphite (HOPG) has been reported \cite{Esquinazi2}.

Graphene, perhaps the mostly researched C-based material, is the two-dimensional (2D) parent for the bulk (3D) material graphite.
The ideal structure consists of C atoms arranged in a hexagonal, honeycomb pattern with parallel graphene layers 
at a (lattice constant) $\bar{c} = 3.35$ $\textnormal{\AA}$ distance apart and weakly interacting by Van der Waals forces. 
In-plane, each C atom is covalently bonded to three other C atoms at a (the other lattice constant) distance 
$\bar{a} = 1.42$ $\textnormal{\AA}$ through $sp^{2} - sp^{2}$ axial hybrid orbital overlap.
This layered structure of graphite results in highly anisotropic physical properties. 
In addition, ion-implantation, heat treatment, hydrogenation, or oxidation result 
in $sp^{2}$ to $sp^{3}$ bond conversion. Distorted $sp^{2}$ C bonds form grain boundaries \cite{Koch}.
Graphite is also the base material for C fibers. Hybrid C fiber-HTS materials are used as stronger, flexible, 
and chemically stable HTS wires for SC magnets, particle accelerators, NMR devices 
or electromagnetic interference shielding covers for spacecrafts. NbN-coated C fibers \cite{Pike} 
have critical densities $J_{c}$ of the order of $10^{6}$ A/cm$^{2}$ and 
critical fields $B_{c2}$(0) up to 25 T \cite{Dietrich}. Improving flux pinning in HTS materials is also very important \cite{Haugan} 
and needs to be considered for each generation of new HTS wires.

The fundamental physical processes leading to the observed temperature($T$)-dependent transport properties of disordered systems such as C fibers 
have been reviewed \cite{Brandt1,Klein,Wang}.
Due to the inherent disorder, the electrons are in general confined to particular regions of the lattice, a phenomenon known as localization. 
The random potential acts as a trap for electrons, which become localized within the region of the trap. 
Energetically, the hopping would rather occur to an energy level close to the one for a neighboring state. The trapping results in inelastic
scattering in the electron-electron interactions. 
The $T$-dependent electrical conductivity is described by Mott law:
\begin{equation}
\sigma(T) = \sigma_{0}\textnormal{exp}[{-(T_{M}/T)^{1/(d+1)}]}
\end{equation}
where $T_{M}$ is the Mott temperature and the power factor $d$ takes integer values 0 to 3 depending 
on the dimensionality of the electronic transport \cite{Mott}. 
At the same time, the trapped electrons can interact and form Cooper pairs. In SC materials, disorder can play an intricate role. 
Trapped charges can suppress the coherence of phase slips, thus favoring SC correlations \cite{Bard}. 
An increase in the 2D density of charges through material intercalation can enhance 
the likelihood for the occurrence of quantum phenomena such as 2D SC \cite{Kulbachinskii}. 

In this work, we are reporting resistivity, magnetoresistance, nonlocal electric differential conductance, 
and DC magnetization measurements conducted on octane-intercalated C fibers.
In addition, we have included results on octane-intercalated graphite powder as well as a HOPG-insulator(Kapton tape)-HOPG composite sample.
Below $T \sim 50$ K, the magnetoresistance data shows a transition from AF to FM correlations as the 
strength of the magnetic field is increased. In addition, the irreversibility observed in the field-cooled vs. the zero-field cooled data 
for a sufficiently high magnetic field suggests that the system also enters a SC state below $T_{c} \sim 50$ K. These results are also corroborated
with the energy gap data obtained from measurements. 
An excitonic mechanism is likely driving the system to a SC state below the same $T \sim 50$ K, where the gap is divergent.
The octane-intercalated C fiber is a multiple gap system, thus pseudogapped, with critical temperatures estimates above room temperature.
The temperature dependence of the SC gap follows the flat-band energy relationship. 
We find that the flat band gap parameter linearly increases with the temperature above $T_{c} \sim 50$ K.
Thus, either a magnetic or an electric field can drive this hydrogenated graphitic system to a SC state below $T_{c} \sim 50$ K.
We also find evidence of pseudogap states above $T_{c} \sim 50$ K, which are likely due to AF spin fluctuations.

\section{Experiment}
The C fibers used in this study are the polyacrylonitrile ((CH$_{2}$-CH-CN)$_{n}$) or PAN-based of T300 type with a C content $\sim93\%$.
C fibers are usually subjected to graphitization at $T$ up to 3000 $^{0}$C \cite{Ekin}. 
In particular, the T300 C fibers are heat-treated to $T_{HT} = 1500$ $^{0}$C thus resulting in a mass density of $\rho_{m} = 1.8$ g/cm$^{3}$. 
The fibers are turbostratic, with volumes of parallel nearest-neighbor C layers randomly rotated such that the overall structure 
looks random on the smaller scale while quasi-1D on the larger scale. 
As a semimetal, graphite has both kind of charge carriers, electrons and holes, contributing
to the electric conduction. At input currents low enough to not significantly affect the geometry of the fiber, 
the $T$-dependence of the electrical resistivity $\rho(T)$ is mainly determined by the
electronic transport properties of the charge carriers, their densities, and their mobilities. 
Pristine T300 C fibers are strongly $n$-type, i.e., having electrons as dominant charge carriers.
At $T = 300$ K, $\rho \simeq 1.8$ m$\Omega\cdot$cm vs. $\simeq 1.7$ $\mu\Omega \cdot$cm for Cu. 
The C fibers were subjected to intercalation with $99.99\%$ purity octane during a three-day time interval. 
While a term like 'soaking' can be used to explicitly describe the treatment of graphite with alkanes, 
the octane intercalation leads to the formation of hydrogen-rich puddles in between neighboring graphitic layers.
The quality of the electrical contacts was optically checked using an Olympus BX51
microscope equipped with a digital camera (Fig. \ref{Fig1}a). 

Temperature-dependent resistivity $\rho(T)$ measurements without magnetic field 
were carried out using a Gifford-McMahon cryocooler and a LakeShore 340  Controller.
The electrical current was sourced through a Keithley 2430 1 kW PULSE current-source meter and the electrical potential difference (voltage) 
along the sample was measured with a Keithley 2183A Nanovoltmeter.  The vacuum was controlled by 
a Laser Analytics TCR compressor and a Pfeiffer turbo pump. 
The sapphire substrate used had four $\sim$2 mm wide silver (Ag) strips affixed via C dots. 
The ratio of the gap between the current leads and the and gap between the voltage leads was close to four,
as required by the four-wire Van Der Pauw technique \cite{Ekin}.
The C fiber was affixed perpendicularly to the Ag strips by dot-like contacts using colloidal Ag. Cryogenic grease \cite{LakeShore} was used to assure 
good thermal contact between the C fiber and sapphire substrate. With the C fiber's thermal conductivity
about ten times larger than of the grease and about twenty times larger than of the sapphire substrate \cite{Ekin},
the most significant heat transfer belonged to the C fiber. The sample was placed on the aluminum heater block
and four POGO pins \cite{Everett} spring-pressed on the C fiber completed the measurement circuit. 
In order to avoid self-heating effects, the power level was maintained low by sourcing a current $I \sim$ $\mu$A through the C fibers. 
Both the few mm length and the cross-section of the $\sim 7$ $\mu$m (average) diameter C fiber modify with  $T$ in an anisotropic fashion that is 
quantified by the in-plane and out-of-plane $T$ coefficient for $\rho$, $\alpha_{a} \cong -3.85 \times10^{-6}/^{0}$C and 
$\alpha_{c} \cong 9 \times10^{-6}/^{0}$C, respectively. Nevertheless, the $T$-change of the ratio between the cross-sectional area and the length 
of the C fiber was very small and thus neglected. When immersed in the cryogenic fluid, the $T$ difference 
between the axis and the surface of the C fiber is given by $\Delta T=RI^{2}/(4\pi K_{T}l)$. The transverse (along the $c$ axis) thermal 
conductivity is $K_{T} = 10^{-2}$ W$\cdot$m$^{-1}\cdot$K$^{-1}$. When a few-millimeter long C fiber is sourced by a current in the 
$\mu$A range, $\Delta T \sim10^{-2}$ K. Each measurement was taken with the C fiber $T$-stabilized by the surrounding cryogenic fluid. 
Magnetoresistance measurements down to $T = 1.9$ K were carried out using the 6500 Quantum Design of the Physical Properties Measurement System \cite{QD}.

\section{Results and Discussion}
\begin{figure}
\centering
\includegraphics[width=3.2in]{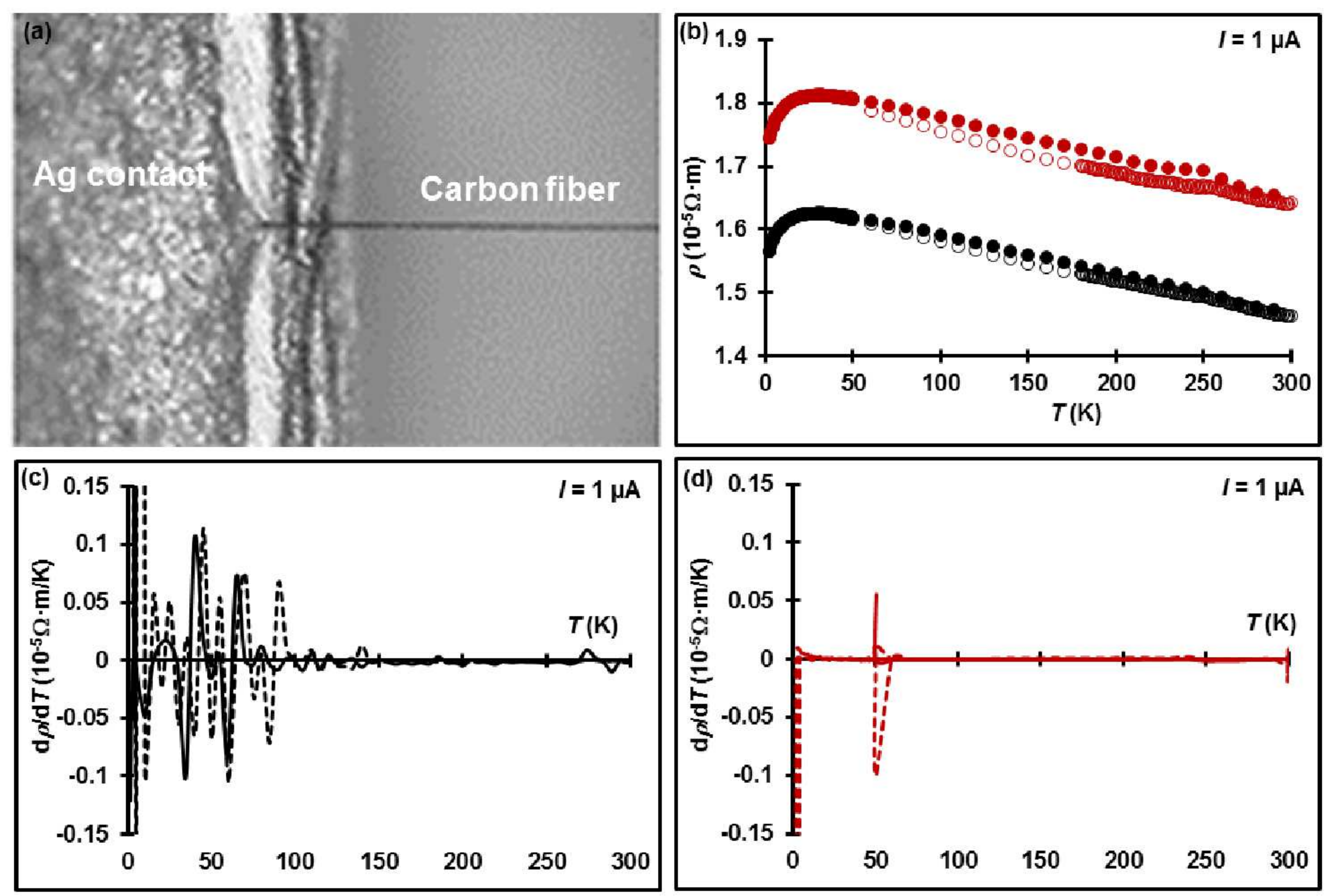}
\caption{Temperature-dependence of the electrical resistivity $\rho(T)$ and its derivative $d\rho(T)/dT$ for a raw (in black) 
and an octane-intercalated C fiber (in red), respectively. 
The inset shows the optical microscopy image of the Ag-metal current contact to the C fiber sample. 
The cooling (empty symbols) and warming (filled symbols) data was taken using a small direct current $I = 1$ $\mu$A.}
\label{Fig1}
\end{figure}

The $T$-dependent electrical resistivity $\rho(T)$ is shown in Fig. \ref{Fig1}b. The input current was kept small, $I = 1$ $\mu$m.
Both the raw and the octane-intercalated C fiber show a well-localized insulator-metal-insulator (I-M-I) transition at $T \sim 250$ K and a 
decrease in $\rho$ below $T \sim 50$ K.
The $T$-derivative of the resistivity $d\rho(T)/dT$ (Fig. \ref{Fig1} c and d) shows a possible transition below $T \sim 50$ K. 
In addition, the larger hysteresis observed between the cooling 
and the warming data observed with the octane-intercalated C fiber suggests irreversible behavior.
\begin{figure}
\centering
\includegraphics[width=3.2in]{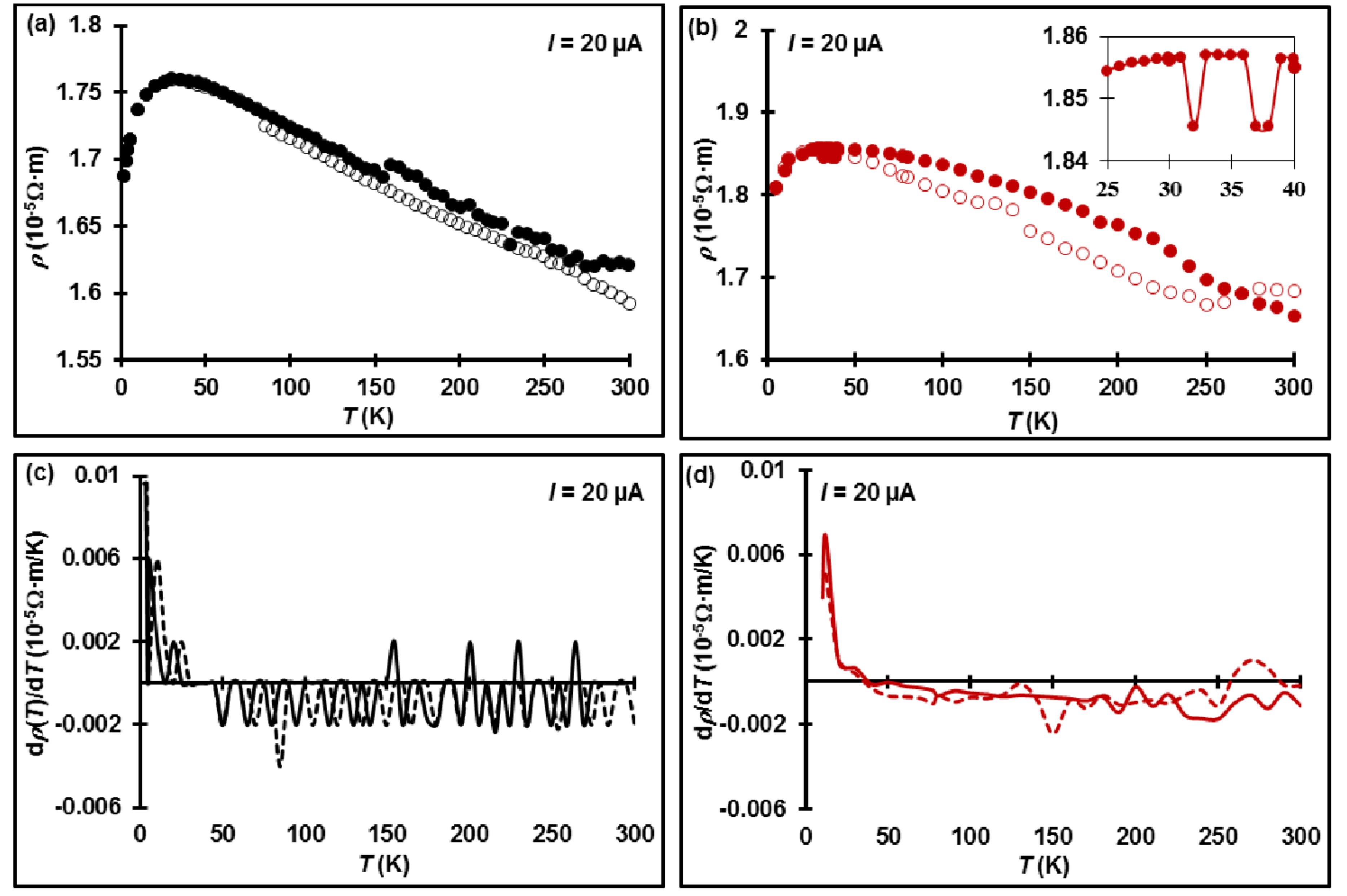}
\caption{Temperature-dependence of the electrical resistivity $\rho(T)$ and its derivative $d\rho(T)/dT$ for a raw (in black) 
and an octane-intercalated C fiber (in red), respectively. The cooling (empty symbols) and warming (filled symbols) data 
was taken using a direct current $I = 20$ $\mu$A.}
\label{Fig2}
\end{figure}

Searching for possible SC-like behavior in C fibers, any drop in $\rho$ provides impetus for further investigation.
A significant M-I-M transition occurs at $T \sim25$ K where $\rho$ reaches a maximum.
Similar $\rho(T)$ behavior was found before in PAN-based C fibers after significant heat treatment/carbonization \cite{Dejev,Spain},
as well as in exfoliated graphites \cite{Uher1}.
Anomalous states have been observed also in early HTS materials, the hole-doped perovskites LaCu$_{2}$O$_{4}$ \cite{Bednorz,Chu2}.
The presence of I-M transitions reflects the $T$-dependent balance of charge carriers'
densities and mobilities. Unless a following I-M transition occurs, for the SC to set in the electronic instability leading to a M-I transition needs to be avoided.
The origin of the M-I transition has been debated and several explanations have been proposed: 
a) magnetic freeze-out of impurity carriers, which is attributed to the magnetically-induced localization of impurity-type 
carriers on charge centers \cite{Brandt2};
b) Wigner crystallization of the free electron gas at low $T$ to a non-metallic and AFM state \cite{Wigner};
c) Anderson localization \cite{Anderson1} of charges in certain random fields such as 
the ones found in disordered systems like C fibers. Below $T \sim25$ K, $\rho(T)$ for these C fibers has an obvious metallic-like
behavior, which appears to dominate over the background electrical conduction for a disordered system, the latter being
described within the variable-range hopping mechanism. At low $T$, the hopping contribution should vanish, leaving out only 
a metallic dependence $\rho(T) = \rho_{0}(1 + \alpha T)$. At very low values $T \leq 1$ K and no
magnetic field, it was found that the conductivity $\sigma (= 1/\rho)$ follows the empirical $T$-dependence:
\begin{equation}
\sigma(T) = \alpha + \beta \textnormal{ln}(T/T_{1}) + \gamma(T/T_{2})^{-1/2} + \delta(T/T_{3})^{1/2},
\label{Koike_eq}
\end{equation}
\noindent
where the ln$T$ term (for a clean system) and the quantum correction $T^{-1/2}$ (for more disordered systems) quantify, respectively, 
the Kondo effect owning to the presence of
localized spins in the heat-treated C fibers \cite{Koike}. The $T^{1/2}$ term is needed for the case when magnetic fluctuations arising for
instance from the application of high magnetic fields would lead to the suppression of the Kondo effect and in turn strong
electron-electron interactions become important. In the absence of a magnetic field, the sign of $\gamma$ is negative, corresponding 
to the AFM exchange interaction $J_{exch} < 0$.

Additional features of the electronic transport in octane-intercalated C fibers are observed at larger input current, $I = 20$ $\mu$A (Fig. \ref{Fig2}).
\begin{figure*}
\centering
\includegraphics[width=6.5in]{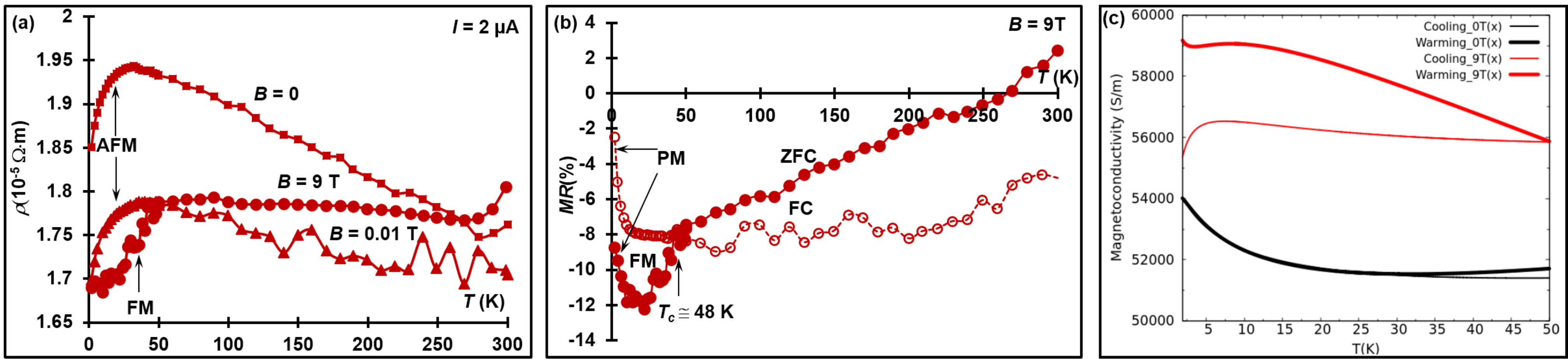}
\caption{a) Magnetic field effect on the temperature-dependent resistivity for octane-intercalated C fiber. 
b) Magnetoresistance data for the magnetic induction set to $B = 9$ T. 
The zero-field cooled (ZFC) line goes below the field-cooled (FC) line at $T \simeq 48$ K.
c) The fit of the magnetoconductivity data for zero magnetic field (in black) and for a magnetic field $B = 9$ T (in red) 
to Eq. \ref{Koike_eq} below $T = 50$ K. The zero field data shows AFM during both cooling and warming, while the data taken for a field 
$B = 9$ T shows AFM ($\gamma > 0$) during cooling (FC) and FM ($\gamma < 0$) during warming (ZFC).}
\label{Fig3}
\end{figure*}
At $T \simeq 40$ K, the irreversibility between the cooling and the warming data practically disappears and several metastable states are observed.
Notice the closeness of the data point at $T = 40$ K to the $T_{c} = 39$ K for MgB$_{2}$ could be more than just a mere coincidence. MgB$_{2}$ is considered
both crystallographically and electronically equivalent to nonstaggered graphite (the B$^{-}$ layer) that has undergone a zero-gap semiconductor-SC
phase transition by large $c$-axis chemical pressure due to Mg$^{++}$ layers \cite{Baskaran1}. 
The electronic similarities between graphite and other systems like MgB$_{2}$ or $\textnormal{Bi}_{2}\textnormal{Se}_{3}$ are 
described within the resonating-valence-bond (RVB) model. The RVB states are preexisting Cooper pairs in the potential zero-gap semiconducting
B$^{-}$ sheets that manifest themselves as a high-$T_{c}$ SC ground state \cite{Baskaran2,Baskaran1}.
RVB can also explain SC below $T_{c} = 19.3$ K in K$_{3}$C$_{60}$ \cite{Pauling}, as well as 
anomalies like the 41 meV peak observed in the phonon density of states for YBCO \cite{Liarokapis,Arai}.

The effect of a perpendicular magnetic field on the transport properties of the fiber 
was tested through measurements on the specific resistance also known as magnetoresistance: $MR(\%) = 100\times[R(B) - R(B = 0)]/R(B = 0)$.
One noticeable feature is the decrease of $\rho$ with increasing the field strength (Fig. \ref{Fig3}a).
The transverse magnetic field applies a magnetic pressure on the fiber, enhancing the anisotropy in the electronic transport.
At $B = 9$ T, the magnetic pressure $p_{mag} = B^{2}/2\mu \simeq 32$ GPa.
The fiber becomes more metallic, thus $\rho$ decreases with increasing the strength of the magnetic field. 
The irreversibility (hysteresis) in the high-field magnetoresistance is particularly significant.
Fig. \ref{Fig3}b shows the magnetoresistance corresponding to the largest applied magnetic induction, $B = 9$ T.
Extending the validity of Eq. \ref{Koike_eq} to the interval $T < 50$ K and doing the fit to the conductivity data (Fig. \ref{Fig3}c), 
the coefficient proportional to the number of spins is found: 
$\gamma_{cooling} \simeq -629$ S/m $< 0$ and $\gamma_{warming} \simeq 720$ S/m $> 0$, respectively.
I.e., for $T < 50$ K the octane-intercalated C fiber shows AFM during its cooling and FM during its warming.
Moreover, the ratio $(\gamma/\delta)(T_{3}/T_{2})^{1/2}$ in Eq. \ref{Koike_eq} goes from $\simeq -2.5$ for the case of no magnetic field
to $\simeq -5.2$ for $B = 9$ T. I.e., the application of a high magnetic field results in the diminishing (to the suppression)
of the Kondo (spin) effect and the promotion instead of electron-electron interactions of possible SC nature.
The magnetic field separates the spin-singlet state from the FM spin-triplet state.
The spin-singlet is observed at the cooling of the system, while the spin-triplet is observed at the warming of the system.

While these are magnetoresistance and not magnetization measurements, Fig. \ref{Fig3}b clearly shows 
that the field-cooled (FC) data goes above the zero-field cooled (ZFC) data below $T \simeq 48$ K.  
A triplet spin state would make possible the coexistence of SC and FM below $T_{Curie} \sim T_{c} \sim 48$ K.
These results suggests that under the application of a high magnetic field, the octane-intercalated C fiber
is likely a FM superconductor (FMSC) below $T \simeq 48$ K.
The effective critical Curie $T_{Curie} \simeq 48$ K is close to the 50 K value found for graphite flake \cite{Barzola1}. 
In addition, as AFM is seen at the cooling of the sample and with the magnetoresistance linearly increasing with $T$, a N\'{e}el transition might also 
occur at $T_{N} \sim 50$ K. Thus, is it possible that $T_{Curie} \sim T_{N} \sim T_{c} \sim 50$ K is a \textit{tricritical} point \cite{Dagotto}.
The magnetic field acts like a chemical potential, which can turn the low-$T$ AFM into FM. Similar to the case of UGe$_{2}$ \cite{Mineev},
here the FM and the FMSC below 50 K might result after two quantum phase transitions. 
Competition between magnons and paramagnons is also possible, i.e., competing pair-making and pair-breaking magnons \cite{Karchev}.
In addition, we observe low-$T$ PM manifested in the upward turn in the magnetoresistance.
This ``reentry'' phenomena was observed in other high-$T_{c}$ systems \cite{Yeshurun} and attributed to the magnetization of the superconductor coming from three
sources: (a) the diamagnetic shielding moment; (b) trapped flux; (c) a PM contribution (positive susceptibility $\chi$) from one elemental component 
(possible the nitrogen, in the case of PAN-derived C fibers).
The Meissner fraction depends strongly on the applied field. In the high-field limit, the trapped flux fraction can be comparable to and eventually can
cancel out the diamagnetic shielding fraction. 
The PM is also a feature of surface SC \cite{Podolyak}.

These results are not entirely surprising. As known, graphite has AFM correlations between unlike sublattices (\textit{ABAB}...) and FM correlations between like sublattices
(\textit{AAA}... or \textit{BBB}...). In graphene \cite{Semenov}, statistically meaningful imbalance in the number of vacancies/defects between
the A and B lattices, i.e. when $|\Delta N_{d}/N_{d}|$ is statistically nonzero, leads to the nucleation of ferrimagnetic puddles and their
subsequent growth into sizable FM domains. In graphene, zigzag edges can be either FM, AFM, or ferrimagnetic.
Magnetic order and SC have been also observed in bundles of double-wall CNTs \cite{Barzola2}.
Following a BCS approach in two dimensions (with anisotropy), a $T_{c} \sim 60$ K has been estimated for a density of electrons per graphene
plane $n \sim 10^{14}$ cm$^{2}$, a density that might be induced by defects and/or H ad-atoms or by Li deposition \cite{Scheike}.
It has been also discussed in \cite{Samatham} how the increased magnetic field changes the helical spin structure to conical structure and eventually 
to fully polarized FM alignment above a critical magnetic field.
In our case, the observed FM is likely due to the octane intercalation, which introduces the spin carrying protons (H$^{+}$ ions).
The resultant magnetic correlations can mediate excitonic correlations. 
The FM instability might also be due to a Lifshitz transition, which has been related to 
the thermoelectric properties of bilayer graphene \cite{Suszalski,Zhang}. 
The Lifshitz transition has been also observed in other systems as due to valence fluctuations that can lead to dramatic changes 
in the Fermi surface topology \cite{Chatterjee} or when it is pushing the system closer to a magnetic instability 
such that the enhanced magnetic fluctuations eventually lead to the reentrance of SC in a FMSC \cite{Sherkunov}. 

\begin{figure*}
\centering
\includegraphics[width=6.5in]{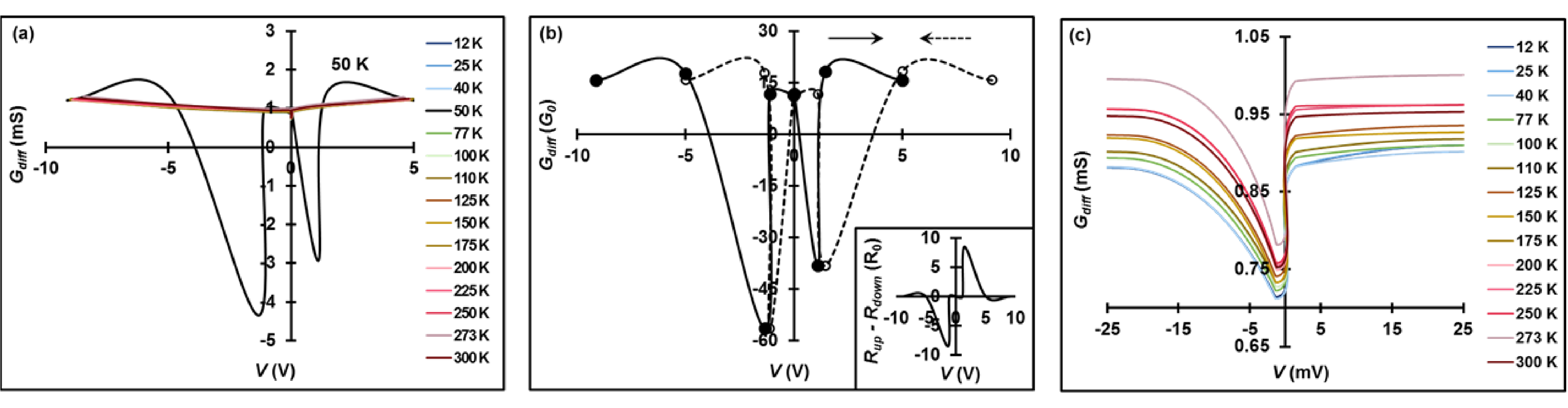}
\caption{Nonlocal $G_{diff}(V)$ data for an octane-intercalated C fiber at high (a) and low voltages (c), respectively. 
b) The gap at $T = 50$ K suggest interference of chiral asymmetric Andreev edge states and crossed Andreev conversion.}
\label{Fig4}
\end{figure*}

\begin{figure}
\centering
\includegraphics[width=3.2in]{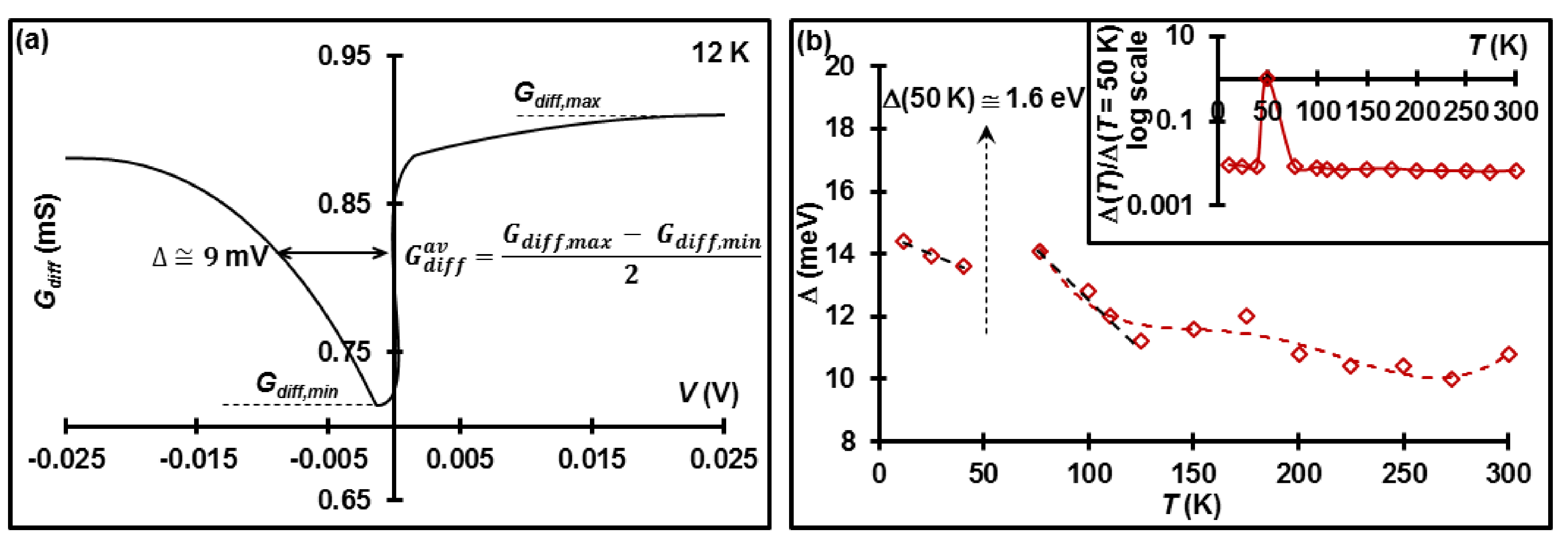}
\caption{a) Example calculation of the gap for the octane-intercalated C fiber at $T = 12$ K. b) $T$-dependence of the energy gap $\Delta(T)$ extracted 
from the nonlocal $G_{diff}$ measurements $G_{diff}(V)$ shown in 
Fig. \ref{Fig4}b. $\Delta(T)$ appears linear around $T \simeq 50$ K and nonlinear at higher $T$.
The inset shows all data, with the gap normalized to the very high value observed at $T = 50$ K.}
\label{Fig5}
\end{figure}

Separated in space and bound by the Coulomb (attractive) interaction, the electrons and holes `crystallize'' into electron-hole pairs pretty much like electric dipoles.
The excitons are observed near M-I transitions. Interestingly, in the superfluid state similar London and Ginzburg-Landau type equations describe both
the electric dipole SC and the electric monopole condensate \cite{Jiang}. 
Excitonic signatures in these octane-intercalated C fibers were probed though measurements of $G_{diff}(V) = dI/dV$ obtained from
current-voltage $V(I)$ measurements on the microscopic length of the C fiber (Fig. \ref{Fig4}a). 
While local $G_{diff}(V_{gate})$ measurements are usually done by scanning tunneling spectroscopy, 
the highly disordered nature of the graphitic system suggests instead the need for nonlocal $G_{diff}$ measurements.
Unlike local electrical conductance spectroscopy, nonlocal electrical differential conductance $G_{diff}$ measurements can distinguish between nontopological zero-energy
modes that are localized around potential inhomogeneities and true Majorana edge modes of the topological phase \cite{Rosdahl}.
Considering normal metal leads connected to a SC, two processes are at the core of nonlocal response in $G_{diff}$: 
1) direct electron transfer between the normal leads;
2) crossed Andreev reflection of an electron from one lead into another lead.  
The height of the peak in $G_{diff}(V)$ is proportional to the nonlocal density of states (DOS) for the C fiber. 
The gap in the energy spectrum is a measure of correlations between electrons, in particular Cooper pair correlations in a SC. 
Notice that in this case the concentration of free charges is brought above the threshold not chemically (by doping), instead, by increasing 
the sourced current. The $T$-dependence of the gap for the low voltage region is shown in Fig. \ref{Fig4}c and the corresponding
energy gap calculation is illustrated in Fig. \ref{Fig5}a.
One remarkable feature is the asymmetry in $G_{diff}(V)$. When the particle-hole and the time reversal
symmetries are violated, the differential tunneling (local) electrical conductivity and the dynamic (nonlocal) $G_{diff}$  
are no more symmetric function of applied voltage $V$ \cite{Shaginyan}. This asymmetry can be observed both in the normal and SC phases of strongly correlated systems. 
As in normal Fermi liquids the particle-hole symmetry 
is not violated, the differential tunneling conductivity (local and non-contact measurement) and the dynamic conductance (non-local and contact-based measurement) 
are symmetric functions of $V$. Thus, the conductivity asymmetry is not observed in conventional metals, especially at low $T$.
In fact, the asymmetry in the tunneling conductance feature is an important sign of the underlying Mott character in doped insulating systems \cite{Baskaran1}
that can show unconventional SC. The asymmetry in $G_{diff}(V)$ mirrors the interference of chiral Andreev
edge states, which is a topological phenomenon. In addition, negative $G_{diff}$ (Fig. \ref{Fig4}b) is a result of nonlocal coherence between electron
and holes in the Andreev edge states. Thus, the non-local negative conductance/resistance indicates the presence of
crossed Andreev converted holes, namely the coupling between two quantum Hall edge states via a narrow SC link.
Fig. \ref{Fig4}b shows that the hump spreads from $-\Delta$ to $\Delta$: $2\Delta \simeq (−1.34 + 1.11)$ eV$\simeq 2.5$ eV. This is 400$\times$ the hump width for YBCO.
We are dealing with a chiral spin-triplet $p$-wave with two gap amplitudes:
$\Delta_{V<0} \simeq 1.6$ eV located at $V \simeq -1.34$ eV and $\Delta_{V>0} \simeq 0.6$ eV located at $V \simeq 1.11$ eV.
For voltages smaller than the SC gap at this temperature (50 K), the transport is
dominated by the Andreev reflection, were an incoming electron is converted into a reflected hole.
The gap asymmetry is due to the charge imbalance that create different rates at which
the electron-like and hole-like quasiparticles are evacuated from the Andreev
bound states and/or by the destruction of chirality symmetry by the magnetic
exchange field due to the itinerant FM introduced in the system by octane with its freely moving H+ (protons) on the graphite’s interfaces.
We also notice the Fano-line shape of $G_{diff}$. The Fano line is a manifestation of coexisting polarons and Fermi
particles in a superlattice of quantum wires. 
In the stripes scenario for HTS, a Fermi liquid coexists with an incommensurate 1D charge density wave (CDW) forming a multi-gap SC near a
Lifshitz transition where $T_{c}$ amplification is driven by Fano resonances involving different condensates.
Thus, $G_{diff}$ data shows that SC correlations might be established in the octane-intercalated C fibers below $T \sim 50$ K.
This value is close to the mean-field $T$ for SC correlations in the metallic-H multilayer graphene or in HOPG, 
$T_{c} \sim 60$ K \cite{Garcia2}. 

In order to better understand the nature of SC in the octane-intercalated C fiber samples, we have analyzed 
the $T$-dependence of the full gap $\Delta(T)$, more often denoted by $2\Delta(T)$ (inset in Fig. \ref{Fig5}b).
A very significant peak is observed at $T = 50$ K. 
The $T$-dependence of the gap $\Delta(T)$ was further analyzed with the data point at $T = 50$ K excluded (Fig. \ref{Fig5}b).
As expected, $\Delta(T)$ decreases with $T$. 
Below and above $T = 50$ K, $\Delta$ appears to decrease linearly with $T$. 
This $\Delta(T)$ dependence is not the one expected for a BCS SC, i.e., $\Delta(T) \approx 1.74\Delta(0)\sqrt{1 - (T/T_{c})^{2})}.$
If the gap were a semiconducting one, i.e. $\Delta \equiv E_{g}$, its $T$-dependence would have been like 
$E_{g} \sim k_{B}T^{2}$ for $T \ll T_{D}$ and $E_{g} \sim k_{B}T$ for $T \gg T_{D}$, with $T_{D}$ the Debye $T$ \cite{Ravindra}.
The quadratic $T$ dependence is not seen here, while the linear $T$ dependence is also not valid as $T_{D} \simeq 2430$ K for graphite.
Above $T \sim 130-140$ K, Fig. \ref{Fig5}b shows a nonlinear $\Delta(T)$.

\begin{figure}
\centering
\includegraphics[width=3.2 in]{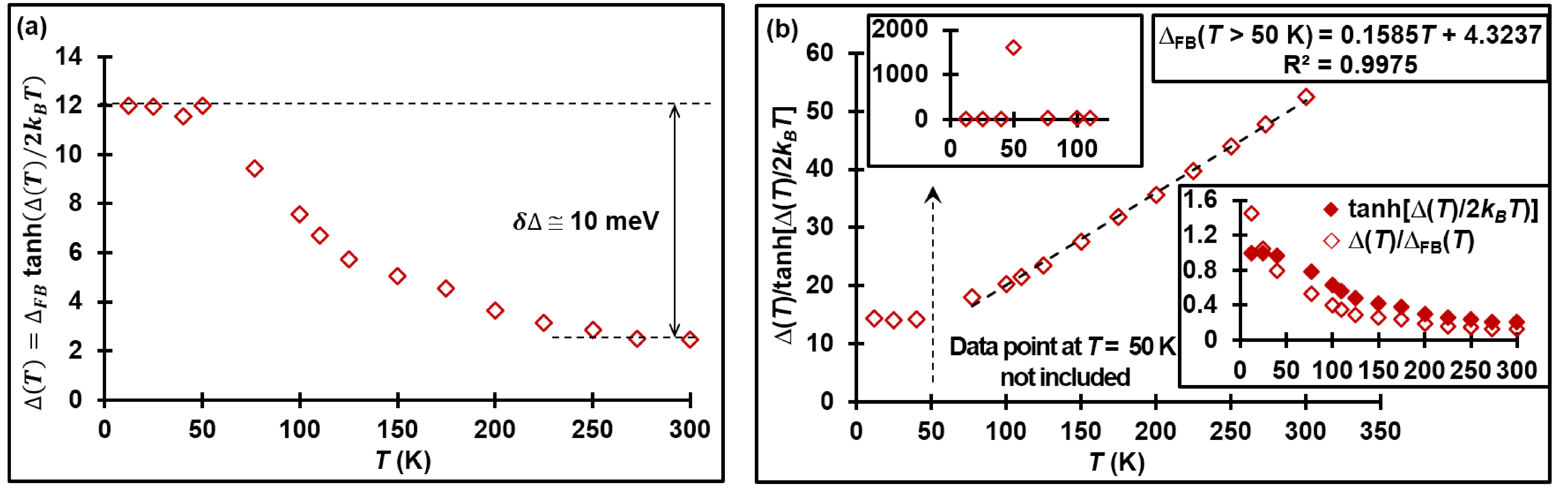}
\caption{a) The flat band $T$-dependence of the gap $\Delta(T)$ with the flat band parameter $\Delta_{FB}$ taken as
the average of all $\Delta(T)$ values except for the data point at $T = 50$ K, $\Delta_{FB} = \Delta_{av} \simeq 12$ meV.
b) The flat band parameter $\Delta_{FB}$ is constant below 50 K and linearly increases with $T$ above 50 K.
The inset shows that a linear $\Delta_{FB}(T > 50 \textnormal{K})$ implies the divergence of $\Delta$ at $T = 50$ K.}
\label{Fig6}
\end{figure}

The $\Delta(T)$-dependence was further analyzed within the framework provided by known results obtained from the application of the mean-field theory.
One unusual SC behavior is manifested in the flat-band (FB) nature of the excitation spectrum, where the 
group velocity goes to zero, $\textnormal{d}\omega(k)/\textnormal{d}k \rightarrow 0$ \cite{Peltonen}. 
When the FBs are partially filled and depending on the filling factor, several ground states can be achieved:
spin-liquid states, quantum anomalous Hall insulators, or chiral $d$-wave SC \cite{Wu2}. 
There is competition between other magnetic states, including AFM and SC.
As the FB states are highly localized around certain spots in the structure, the SC order parameter becomes strongly inhomogeneous. 
The $\Delta(T)$ dependence is the solution to a transcendental equation:
\begin{equation}
\Delta(T) = \Delta_{FB}\textnormal{tanh}[\Delta(T)/2k_{B}T]
\label{FB_gap_eq}
\end{equation}
\noindent
This $\Delta(T)$ dependence comes from the density functional theory approach for SC, where the extended Kohn-Sham equation is written 
in the form of Bogoliubov-deGennes equation used in the conventional theory for the description of inhomogeneous SC \cite{Haruyama,deGennes}.
At the transition, the condition for the FB energy gap $\Delta_{FB}$ translates as $\Delta(T_{c}) \rightarrow 0$.
Taylor series expansion to the first order gives $T_{c} \simeq \Delta_{FB}/2k_{B}$.
Our results for the octane-intercalated C fiber are shown in Fig. \ref{Fig6}a.
The $T$-dependent gap data for the octane-intercalated C fiber $\Delta(T)$ was replotted with the constant $\Delta_{FB}$
taken as the average of all $\Delta(T)$ values except for the data point at $T = 50$ K, 
$\Delta_{FB} = \Delta_{av} \simeq 12$ meV (Fig. \ref{Fig6}a). 
While indeed constant ($T$-independent) below 50 K, the outstanding feature is the linear $T$-dependence of $\Delta_{FB}(T)$ 
for $T > 50$ K (Fig. \ref{Fig6}b). In addition, the lower inset in Fig. \ref{Fig6}b shows that a linear $\Delta_{FB}(T > 50$ {K})
implies the divergence of $\Delta$ at $T = 50$ K.

\begin{figure}
\includegraphics[width=3.2in]{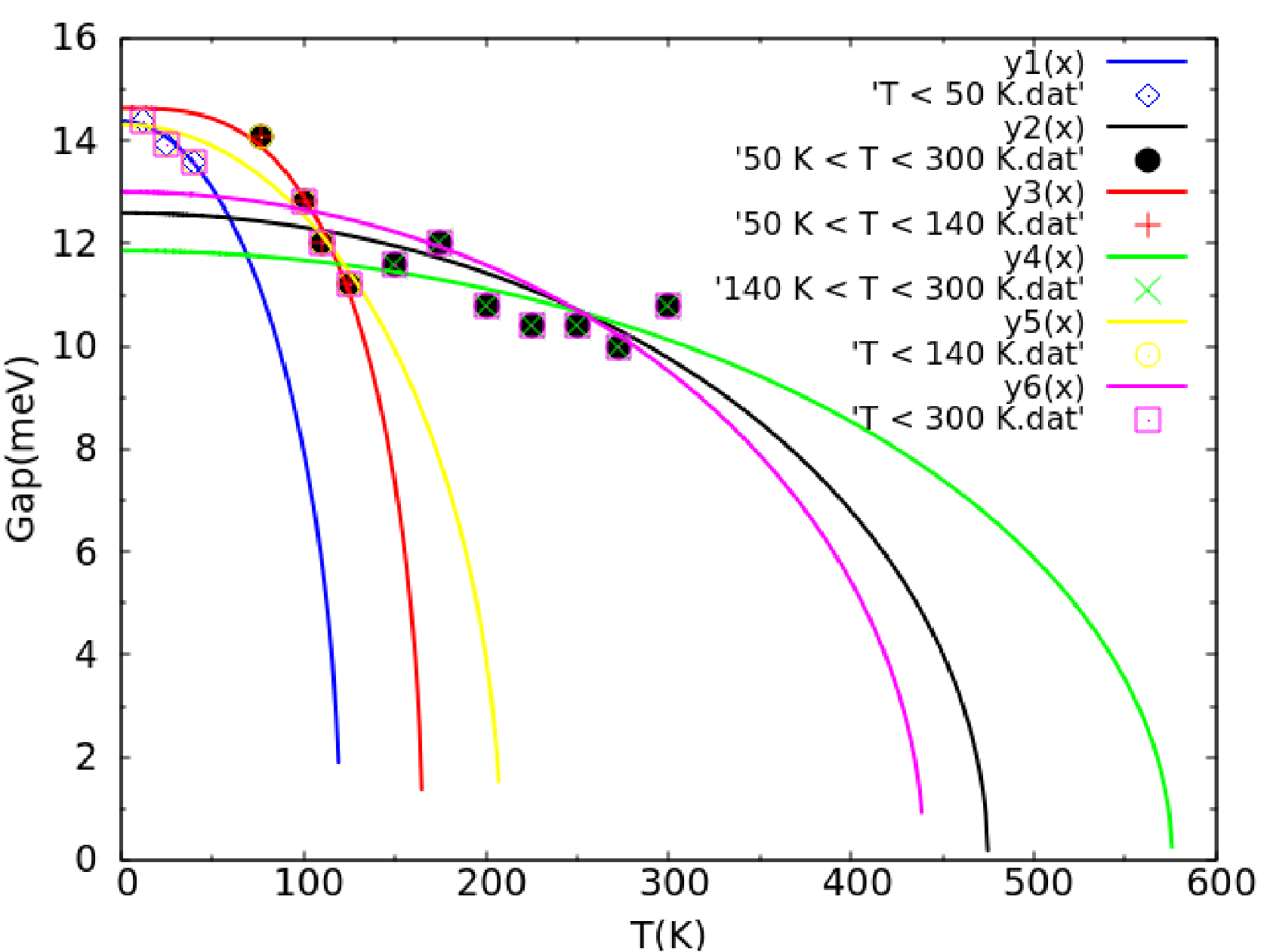}
\caption{Experimental $\Delta(T)$ and the fit to the BCS formula for different $T$ intervals. The fitting parameters are shown in Tab. \ref{Gap_T_BCS}.}
\label{Fig7}
\end{figure}

Fitting the gap data to the BCS dependence for different $T$ intervals (Fig. \ref{Fig7}) gives high $T_{c}$ values, even past room-$T$ (Tab. \ref{Gap_T_BCS}). 
Also shown are the calculated $r_{BCS} = \Delta(0)/k_{B}T_{c}$ ratios, which are significantly smaller than known BCS values. 
The fitting curves for data below and above 50 K, respectively, intersect at $T \simeq 58$ K.
When the delimiting $T$ is 140 K, the fitting curves for data below and above $T = 140$ K intersect at $T \simeq 120$ K. 
The fitting curves for high-$T$ data intersect at $T \simeq 250$ K.
I.e., the delimiting $T$ is close to the one for the intersection of fitting curves for data below and above it, which can be taken as a consistency criterion.
The $r_{BCS}$ values for the octane-intercalated C fiber are lower even than the values predicted in \cite{Margine}, 
thus suggesting that the unconventional SC-like octane-intercalated C fibers have a dominantly FEL energy spectrum.
By comparison, for the phonon-mediated SC MgB$_{2}$, which is crystallographically and electronically similar to graphite, 
the known values for $\Delta \simeq 9$ meV and for $T_{c} \simeq 39$ K give $r_{BCS} \simeq 1.4$.
Likely more than just coincidentally, the same value $r_{BCS} = 1.4$ is obtained by fitting the data for the octane-intercalated C fiber below $T = 50$ K 
(Fig. \ref{Fig6}b and Tab. \ref{Gap_T_BCS}).

Thus, the $\Delta(T)$ fit to the BCS formula found that the octane-intercalated C fiber is a multiple-gap system.
The similar MgB$_{2}$ has also a multiple-gap structure \cite{Takasaki}: a larger gap $\Delta_{\sigma}$ in the 
$\sigma$-orbital band and a smaller gap $\Delta_{\pi}$ in the $\pi$-orbital band. The gap ratios for MgB$_{2}$ are $r_{BCS}^{\sigma} \simeq 4.5$ and  
$r_{BCS}^{\pi} \simeq 1.7$, such that, $r_{BCS}^{\sigma}$/$r_{BCS}^{\pi} \simeq 2.7$. Both energy gaps have $s$-wave symmetry. 
The larger gap is highly anisotropic, while the smaller one is either isotropic or slightly anisotropic. 
Just as in MgB$_{2}$, covalent bonds are driving metallic behavior in graphite as well. 
The similarities in the calculated multiple-gap structure for $n$-doped graphene and MgB$_{2}$ \cite{Margine}, 
as well as the transition here occurring at $T \sim 50$ K that is close the the $T_{c} = 39$ K for MgB$_{2}$, 
might be all more than just a mere coincidence with the results obtained on these octane-intercalated C fibers.
Thus, just as MgB$_{2}$, the octane-intercalated C fibers also show a multiple-gap structure. 
Compared to conventional SC materials, where $r_{BCS} \sim 3.53$, the $r_{BCS}$ ratios reported here 
are noticeable smaller. The case of smaller $r_{BCS}$ ratio was discussed before 
as due to the existence of FEL energy bands \cite{Margine}.
On the other hand, multiple-gap energy structure with the $r_{BCS}$ ratios larger than $\sim 3.53$ have been explained within the bisoliton HTS model \cite{Ermakov1}.
The physical reason for the multiple-gap structure in a M-I-SC system, where the SC has an inhomogeneous structure,
was there related to the presence of several local maxima experimentally found in the voltage dependence of the differential conductance $G_{diff}(V)$.
The SC current in the bisoliton HTS model is accompanied by the deformation of energy bands for a one-particle excitation spectrum.
Unlike the tradition model of motion of the whole Fermi sphere, the bisoliton model does not have the deficiencies related to the violation 
of the Pauli principle and the problem of residual resistance \cite{Ermakov2}.
The multiple-gap was found there to be linked to the inhomogeneity of the SC fraction in the M-I-SC sample.
It is possible that a similar situation is found with the octane-intercalated C fibers here that appear to have a granular SC fraction embedded in
an inhomogeneous host, which is a mixture of metallic and insulating domains. 
A larger number of conductive channels will make an embedded domain metallic vs. the smaller number of conduction channels found in an insulating domain.
We also mention that the very high peak in $\Delta$ occurs at $T = 50$ K, which is close to 
the $T_{c}$ found for the B-doped Q-carbon \cite{Bhaumik}.
\begin{table}
\textsc
\bigskip
  \begin{tabular}{|c|c|c|c|}
\hline
  $T$ range (K) &  $\Delta(0)$(meV) & $T_{c}$(K) & $r_{BCS}$ \\
\hline
  $T < 50$ K & 14.4 & 120 & 1.4 \\
  50 K $< T <$ 140 K & 15.5 & 179 & 1.0 \\
  140 K $ < T<$ 300 K & 11.9 & 576 & 0.24 \\
  50 K $< T <$ 300 K & 12.6 & 475 & 0.31 \\
  $T <$ 140 K & 14.3 & 208 & 0.80 \\
  $T < 300$ K & 13.0 & 440 & 0.36\\
\hline
  \end{tabular}
\caption{Results of fitting the experimental data to the BCS formula for the gap$\Delta(T)$.}
\label{Gap_T_BCS}
\end{table}

\begin{figure}
\centering
\includegraphics[width=3.2in]{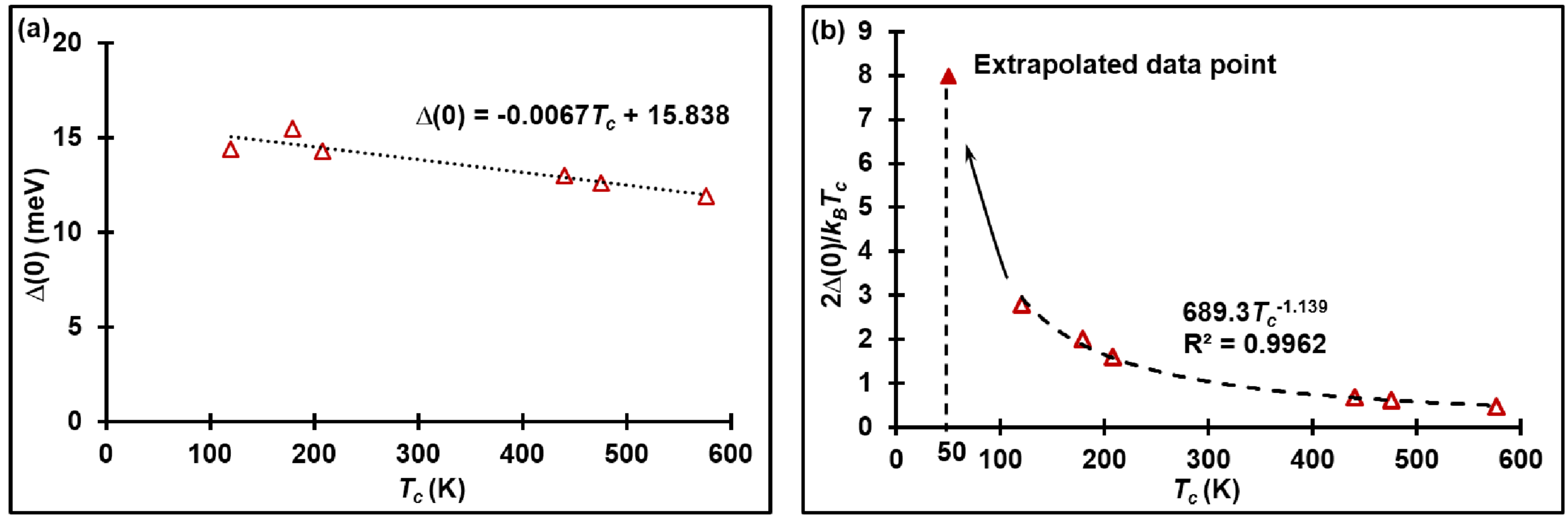}
\caption{a) $\Delta(0)(T_{c})$ dependence. Data from Tab. \ref{Gap_T_BCS}.
b) Power law dependence of the BCS ratio as  a function of $T_{c}$, showing the divergence at $T_{c} \sim 50$ K, where the BCS ratio is $\sim 8$.}
\label{Fig8}
\end{figure}

The high $T_{c}$ for the octane-intercalated C fibers here as found from the BCS dependence of the gap $\Delta(T)$ are close to the most SC-prone 
form of graphite, the one having rhombohedral symmetry. 
In rhombohedral-stacked graphene layers, the in-plane interactions are ferrimagnetic, while the interlayer interactions are AFM \cite{Pamuk}.
The spin polarization opens a gap in the surface state by stabilizing the AFM state.
The $T$-dependent magnetic gap $\Delta_{magn}(T)$ lines when increasing number of rhombohedral graphite layers 
are almost parallel (they do not intersect) except at higher $T$ (close to $\sim 200$ K). I.e., both the $\Delta_{magn}(0)$ gap 
and the N\'{e}el temperature $T_{N}$ increase with the number of graphene layers. Moreover, rhombohedral graphene (\textit{ABC} stacking) displays a conducting surface state with a FB
energy dispersion \cite{Pierucci,Henck}. The electronic ground state in a FB had been predicted to be either SC or magnetic. 
First-principles calculations identify the electronic ground state as an AFM state with a band gap of about 40 meV and a 
$T_{N} \simeq 120$ K. Experiments of mixed \textit{AB} and \textit{ABC}-stacked multilayer graphene are interpreted in terms of SC.
In the case of octane-intercalated C fibers, the estimated $T_{c}$ increases as the upper limit for the $T$-range used for the BCS fit of  $\Delta(T)$ also increases. 
It is possible that the increased number of graphitic layers in the C fiber are contributing 
to the observed behavior, which is also corroborated to the more metallic nature of the heated C fiber. 
The number of channels (located in different graphene layers) contributing 
to the electrical conduction increases with the amount of sourced current.

While SC is generally perceived as a low-$T$ phenomenon, within the FB energy approach it is shown that room-$T$ SC can be achieved
by changing the spectrum and by increasing the DOS close to the Dirac point. In the extreme limit, 
the energy band is the FB surface in the energy spectrum, where the group velocity tends to zero. As a consequence, 
$T_{c}$ is a linear function of the coupling strength $\lambda$: $T_{c} = \lambda\Omega_{FB}/\pi^{2}$, 
where $\Omega_{FB}$ is the FB energy area \cite{Kopnin}. 
Thus, quite high $T_{c}$ can be expected even without extra doping (see references in \cite{Peltonen}).
The electron-phonon coupling strength (or constant) for the octane-intercalated C fibers can be estimated within the FB model as:
$\lambda = a^2\Delta_{FB}/1.3 \times 10^{-3}$ \cite{Peltonen}. Using the averaged lattice constant $\overline{a} = 1.42$ $\textnormal{\AA}$ and 
the average gap $\Delta \simeq 12$ meV, the electron-phonon coupling strength for the octane-intercalated C fibers 
is estimated at $\lambda \simeq 3.0 \times 10^{-38}$J$\cdot$m$^2$
or in the more practical units eVa$^2$, $\lambda \simeq 9.2$ eVa$^2$. This is more than one order of magnitude smaller than the
cutoff value for graphene: $\lambda_{c} = k_{B}T_{D}a^2/1.3 \times 10^{-3} \simeq 140$ eVa$^2$, with the Debye temperature $T_{D} \simeq 2100$ K. 
For graphite, $T_{D} \simeq 2430$ K gives $\lambda_{c} \simeq 160$ eVa$^{2}$.
This large coupling constant, $\lambda > 3$, was also observed in the 2D honeycomb nitride SCs under high-pressure conditions \cite{Hisakabe},
though in that case $\lambda$ was estimated from McMillan's  formula:
\begin{equation}
T_{c} =  \frac{T_{D}}{1.45}\textnormal{exp}\left[-\frac{1.04(1 + \lambda)}{\lambda - \mu^{*}(1 + 0.62\lambda)}\right],
\label{McMillan}
\end{equation}
\noindent
where $\mu^{*}$ is the product of the DOS at the Fermi level $N(0)$ and the Coulomb pseudopotential.
At $T = 50$ K, a very high value $\lambda \simeq 1230$ eVa$^2$ is reached in the octane-intercalated C fiber. 
The very large coupling constant, $\lambda \gg \lambda_{c}$ suggests possible formation of a Bose-Einstein exciton condensate,
which is the limiting case for a  BCS SC. One possible explanation for such a large value of $\lambda$ is the topology  
of graphite that acts as an electro-chemical potential \cite{Atanasov}. The mixed structure of SC and normal grains, which act as Josephson 
junctions, is actually a two-phase state where SC can be observed at room-$T$ \cite{Wolf}.
These high $T_{c}$ states, which have been observed in organic matter, are also referred to as type-III or fractional SC.
Thus, Josephson-like normal-SC junctions are formed in the octane-intercalated C fiber.
The transport properties of Josephson-like normal-SC junctions are determined by the competition between 
two different coherent reflection processes: a) the standard Andreev reflection at the interface between the SC and the exciton condensate 
and b) a coherent crossed reflection at the semimetal with the exciton condensate interface that converts electrons from one layer into the other \cite{Bercioux}.
As a consequence of this competition, $G_{diff}$ has minima at voltages of the order $\Gamma$/$|e|$, 
the order parameter for the exciton condensate. Thus, such minima can be seen as a direct hallmark of the existence of a gaped excitonic condensate.
The excitonic condensate does carry a net dipolar current, which expression is similar to the one
for the supercurrent in a SC, $\bf{J} = |e|\rho_{s}(\nabla$ $\Psi$ $- 2|e|\bf{A})$, 
where $\rho_{s}$ is the density of SC (Cooper) pairs, $\Psi$ is the scalar potential, and $\textbf{A}$ is the vector potential \cite{Balatsky}.

The $\Delta(0)(T_{c})$ was plotted in Fig. \ref{Fig8}a.  
Fitting to a line, a zero gap would be attained at a $T_{c} \sim 2360$ K, as Schrieffer predicted for exotic HTS where
magnetism is important \cite{Schrieffer1}.
Interestingly, within the two band gap model used for MgB$_{2}$, the BCS-like gap increases linearly with $T_{c}$, while the gap characteristic to
organic superconductors and alkane-graphite systems decreases linearly with $T_{c}$ \cite{Gonnelli}.
While the $r_{FB}$ values for the octane-intercalated C fiber (Tab. \ref{Gap_T_BCS}) are smaller than the standard BCS ones, 
the reason for this behavior might have to do with the anisotropic nature of the graphitic material \cite{Margine}.
Notice that the ratio $2\Delta(0)/k_{B}T_{c}$ was found to increase with the coefficient $\Gamma$ defined within
a set of quantum-critical models in which the pairing interaction is mediated by a gapless boson
with local susceptibility $\chi(\Omega) =1/|\Omega|^{\Gamma}$ (the $\Gamma$ model) \cite{Chubukov}.
The ratio $2\Delta(0)/k_{B}T_{c}$ has been recently computed numerically for $0 < \Gamma < 2$ within Eliashberg theory and was
found to increase with increasing $\Gamma$. It was argued that the origin of
the increase is the divergence of $2\Delta(0)/k_{B}T_{c}$ at $\Gamma = 3$.
At the divergence, the gap was found to scale with $T_{c}$ as: $2\Delta/T_{c} \approx 4\pi(\frac{1}{1-\Gamma})^{1/3}$.
Indeed, the gap divergence (Fig. \ref{Fig8}b) for the octane-intercalated C fiber is $\Delta(T = 50$ K$) = 1600$ meV
gives $\Gamma \approx 3$. 
Moreover, the ratio $2\Delta(0)/k_{B}T_{c}$ at the divergence is about 8, a value that is characteristic to HTS materials \cite{Drozdov}.

\begin{figure}
\centering
\includegraphics[width=3.2in]{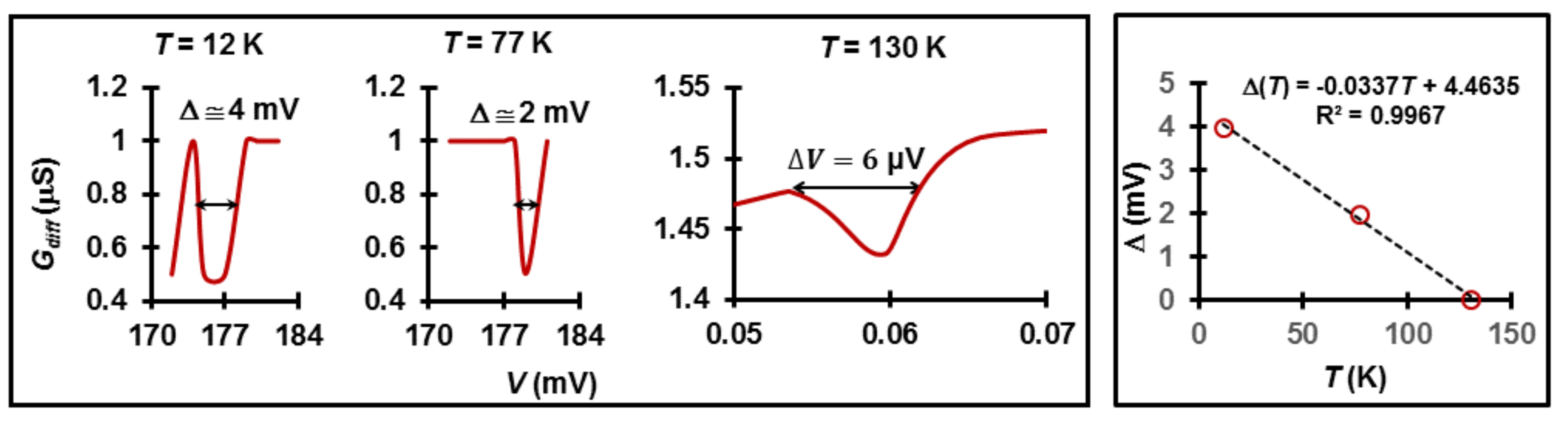}
\caption{a)-c) $G_{diff}$ small gap data at non-zero voltages for $T = 12$ K, $T = 77$ K, and$T = 130$ K, respectively.
d) The $\Delta(T)$ line intersects at $T_{c} \simeq 130$ K.}
\label{Fig9}
\end{figure}

We have also observed gaps in the $G_{diff}$ data at other non-zero voltages (Fig. \ref{Fig9}).
The $\Delta_{S}(T)$ data is almost a perfect line, intersecting the temperature axis at $T_{c} \simeq 130$ K.
The linear $\Delta(T)$ dependence suggests the spin Seebeck effect due to the spin-orbit interaction, where the gap $\Delta$ is a measure of the 
magnetic resonance energy $E_{r} \propto \Delta$ \cite{Yu}. It also points to the excitonic mechanism for unconventional SC based on the 
existence of a spin exciton, i.e. a spin-1 particle–hole excitation.

\begin{figure}
\centering
\includegraphics[width=3.2in]{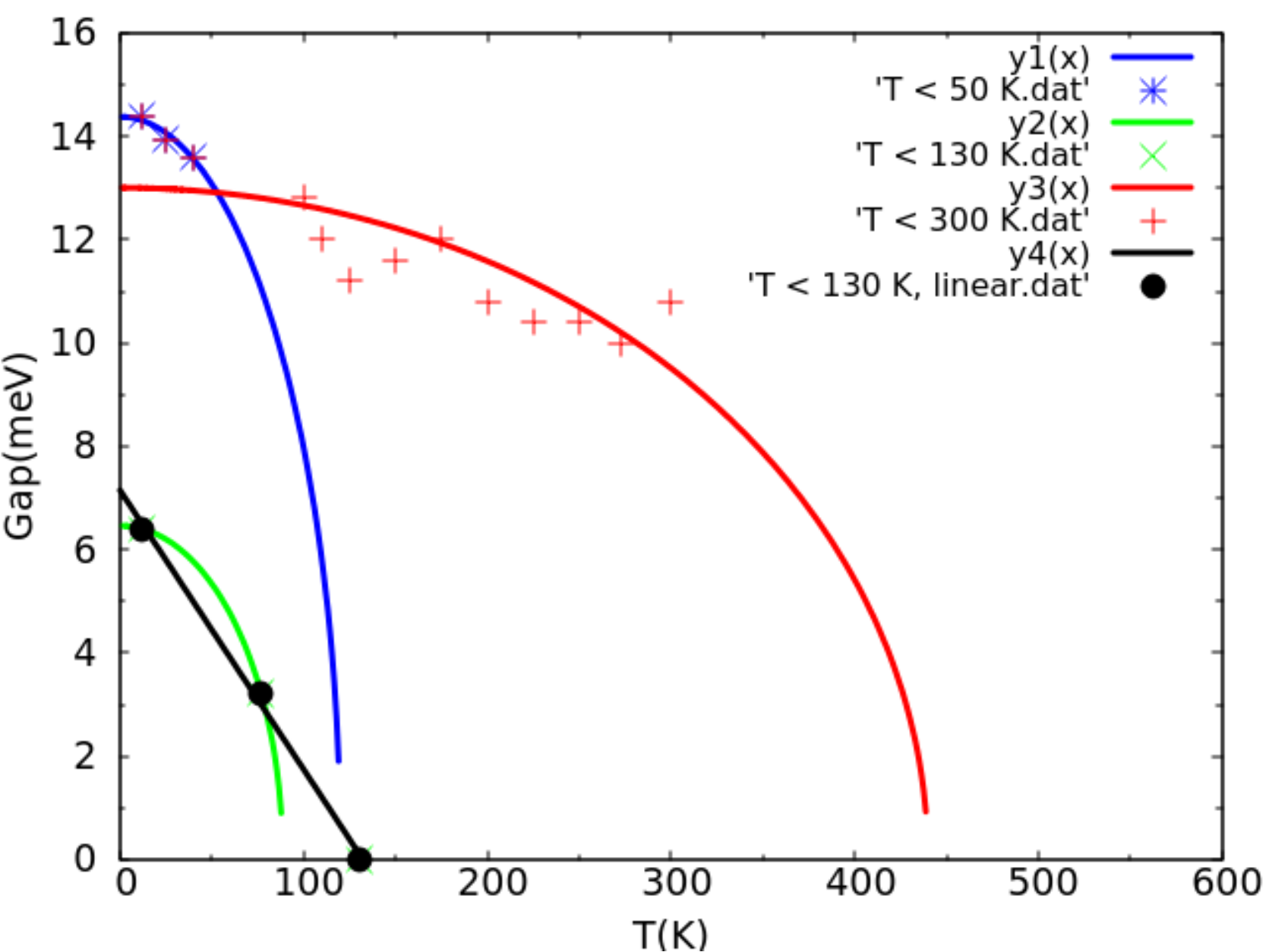}
\caption{Large and small temperature-dependent gap data $\Delta(T)$ for the octane-intercalated C fiber.}
\label{Fig10}
\end{figure}

We have then replotted both the small and the large gap data (Fig. \ref{Fig10}). 
Below $T \sim 130$ K, we observe both a small $\Delta_{S}$ and a large $\Delta_{L}$ gap data. 
While $\Delta_{L}$($T$) (in blue) appears to follow the BCS trend, $\Delta_{S}$($T$) (in black) has the linear spin-wave dependence 
for which the BCS dependence (in green) is not a good fit. Fitting all data below 300 K, the BCS critical temperature is $T_{c} \simeq 440$ K.
We also observe that for the large gap $\Delta_{L} = 13$ meV and the small gap $\Delta_{S} = 6.5$ meV and for $T_{c} \simeq 50$ K,
the gap ratios $2\Delta_{L}/k_{B}T_{c}$ are $\simeq 3.0$ and $\simeq 6.0$, respectively.
Interestingly, these are close to the universality class values found for MgB$_{2}$ \cite{Ekino} and for a other SC materials \cite{Ponomarev}.

\begin{figure}
\centering
\includegraphics[width=3.2in]{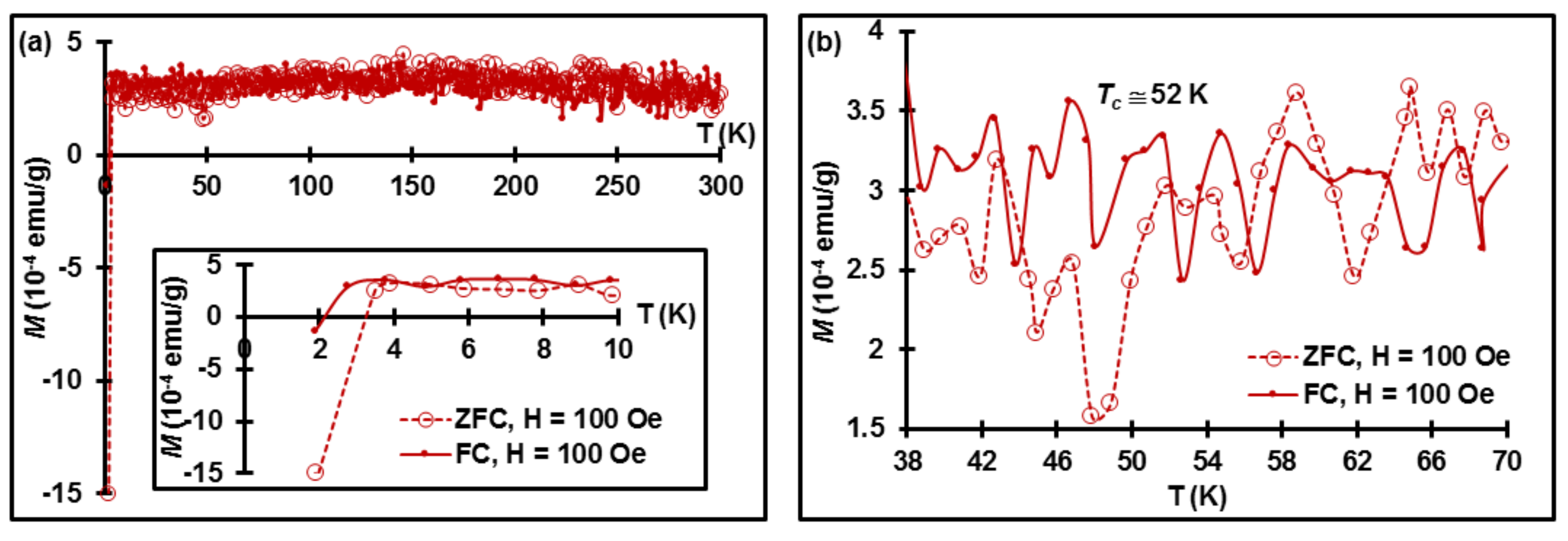}
\caption{a) DC magnetization zero-field cooled (ZFC) and field-cooled (FC) data for octane-intercalated C fiber. 
b) Zoomed data finds an irreversibility temperature $T_{c} \simeq 52$ K.}
\label{Fig11}
\end{figure}

\begin{figure}
\centering
\includegraphics[width=3.2in]{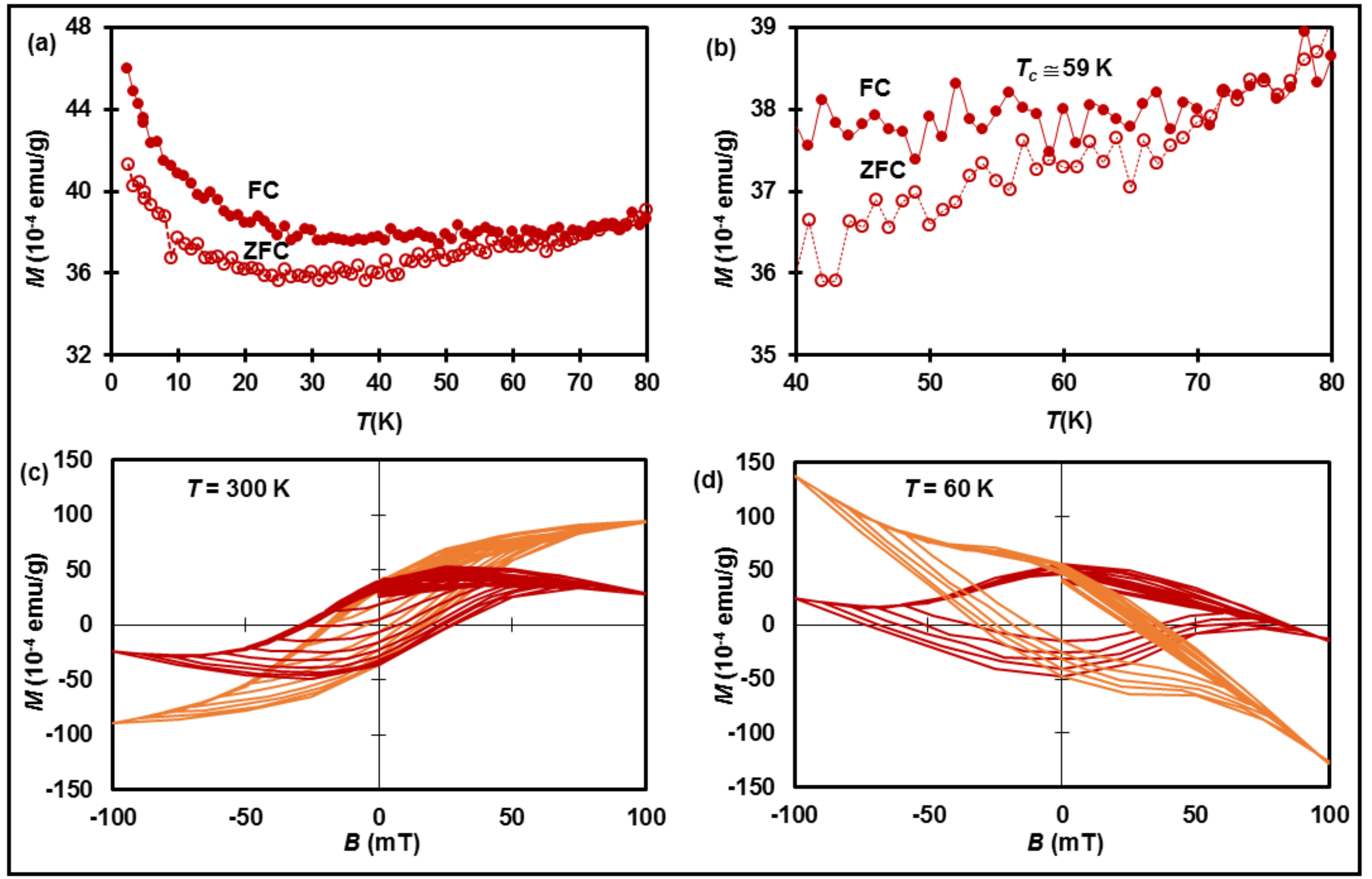}
\caption{a) DC magnetization zero-field cooled (ZFC) and field-cooled (FC) data for octane-intercalated graphite powder. 
b) Zoomed data finds an irreversibility temperature $T_{c} \simeq 59$ K.
c) and d) In-field magnetization data $M(B)$ at $T = 300$ K and $T = 60$ K, respectively, before (in orange) and after (in red) the sample's
diamagnetic background was subtracted.}
\label{Fig12}
\end{figure}

Additional confirmation for finding SC in the octane-intercalated C fibers was provided by DC magnetization measurements using the PPMS 
vibrating sample magnetometry (VSM) option {Fig. \ref{Fig11}}. The irreversibility temperature, where the zero-field cooling (ZFC) 
magnetization goes below the field-cooled (FC) magnetization is $T_{c} \simeq 52$ K. Notably, this $T_{c}$ is close to what has been found before 
from magneto-transport and gap data. We also observe that the magnetization remains positive down to $T \simeq 4$ K, below
which the response becomes diamagnetic. As known, boron-doped diamond is SC below $T_{c} \simeq 4$ K \cite{Ekimov}.
Moreover, octane-intercalated graphite powder from Sri Lanka \cite {RS_mines} shows $T_{c} \simeq 59$ K 
in the ZFC-FC data (Fig. \ref{Fig12}). Significantly, this is close the the mean-field value for $T_{c}$ \cite{Garcia2}.
In addition, our result is also close to what has been found for the as-prepared graphite-sulphur composites \cite{Kopelevich}.

\begin{figure}
\centering
\includegraphics[width=3.2in]{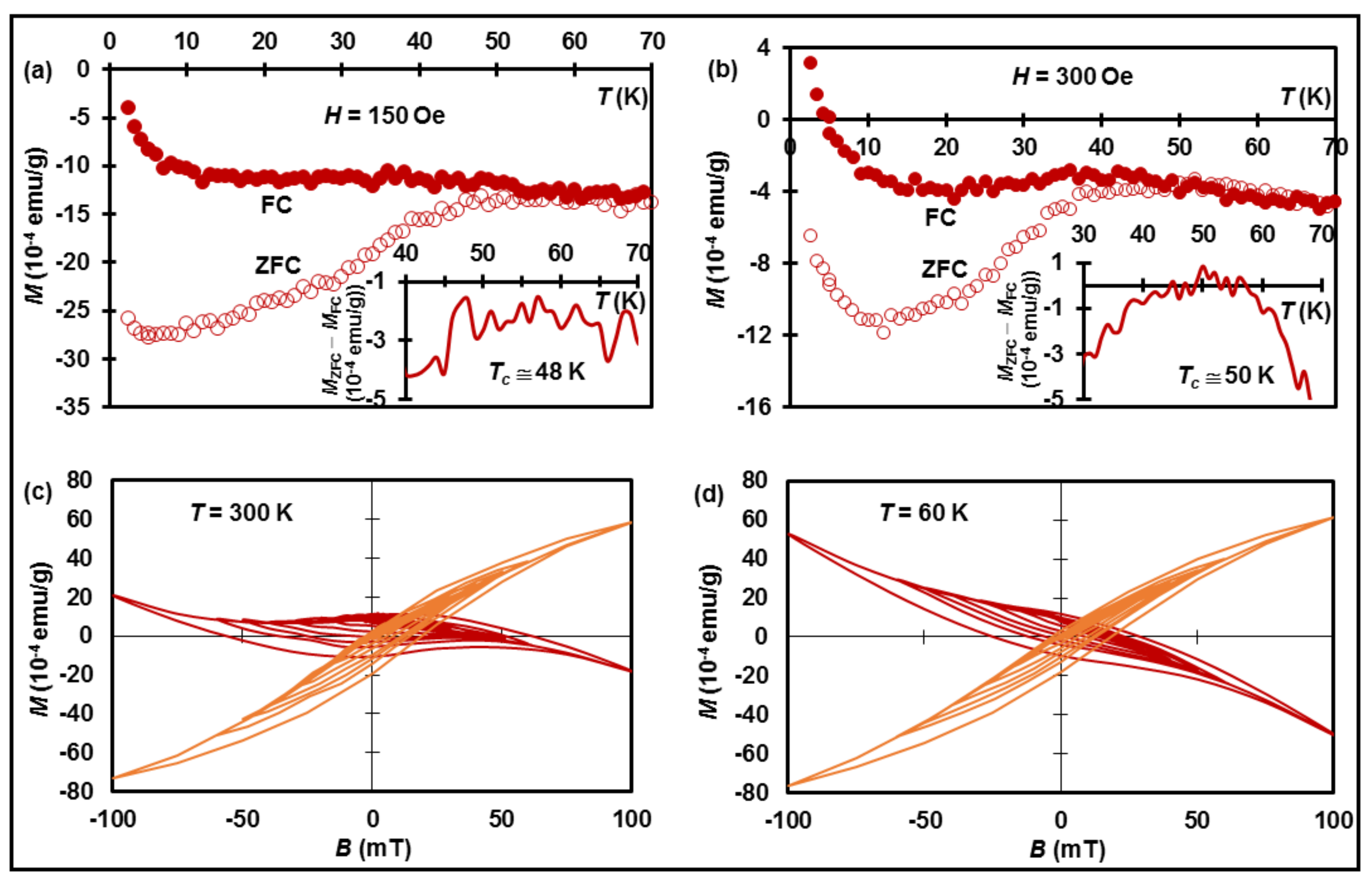}
\caption{a) and b) DC magnetization zero-field cooled (ZFC) and field-cooled (FC) data for a HOPG-Kapton-HOPG composite in $H = 150$ Oe 
and $H = 300$ Oe, respectively. c) and d) In-field magnetization data $M(B)$ at $T = 300$ K and $T = 60$ K, respectively, before (in orange) and after (in red) the sample's
diamagnetic background was subtracted.}
\label{Fig13}
\end{figure}

Following Ginzburg's idea \cite {Ginzburg1}, bulk SC properties were further tested on a composite made out of two Sri Lanka \cite {RS_mines} 
HOPG pieces 800 $\mu$m thick separated by one insulating layer. We have used a $25\mu$m thick Kapton tape \cite{Kapton}, which is often  considered
for cryogenic measurements due to its chemical stability anywhere from 4 K to 673 K. 
The Kapton tape (4,4'-oxydiphenylene-pyromellitimide) contains the same elements as the C fiber (C, H, and N), while double bonded
O is one time present in the Kapton molecule chain plus two O atoms that are connected to each side of the Kapton chain.
The ZFC-FC data for the Kapton tape is shown in Fig. \ref{Fig13}.
Again, the irreversibility temperature  $T_{c} \simeq 48-50$ K was close the all previously found values. Thus, the mean-field predicted $T_{c}$ for graphite
is here experimentally confirmed to be in the 50-60 K range. 

\begin{figure}
\centering
\includegraphics[width=3.2in]{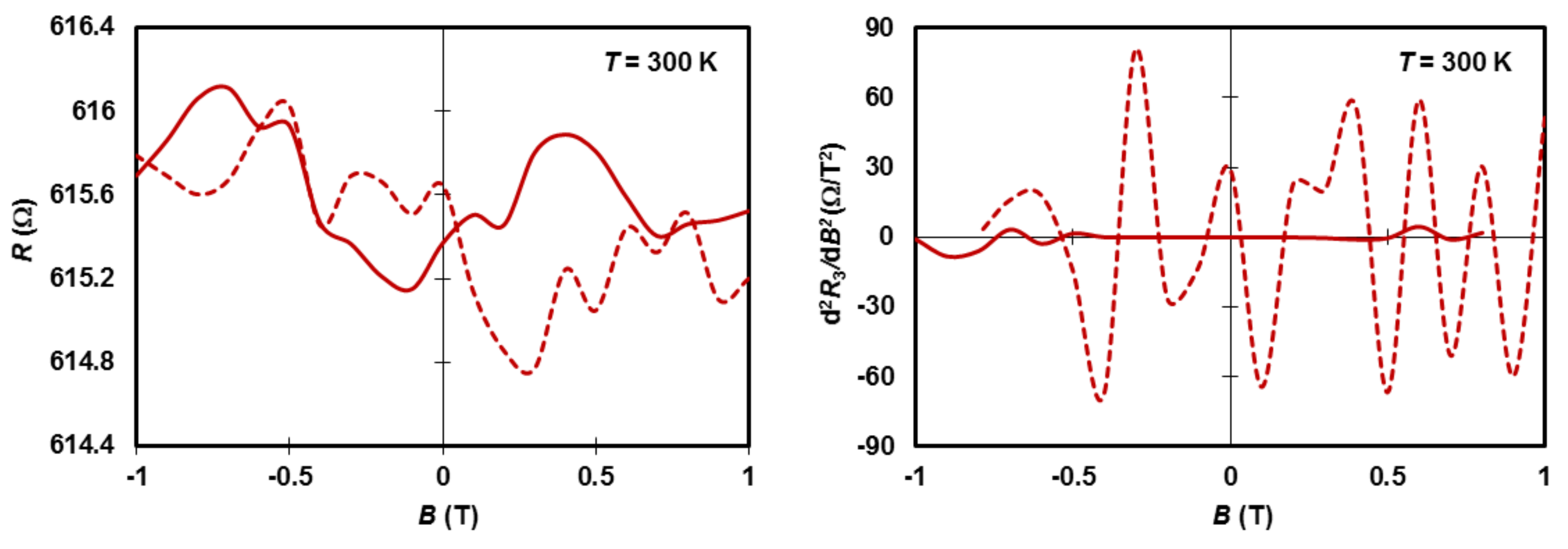}
\caption{Little-Parks (L-P) resistance oscillations at $T = 300$ K in the octane-intercalated C fiber for magnetic fields applied normal to the C fiber at  
decreasing (dotted line) and increasing (continuous line) the field.}
\label{Fig14}
\end{figure}

We have also found Little-Parks (L-P) oscillations in the octane-intercalated C fiber, proving that the C fiber is, indeed, a topological object 
(Fig. \ref{Fig14}. The Little-Parks effect is observed as a modulation of the electrical resistance by a magnetic field applied
perpendicularly to the graphite layers \cite{L-P,Tinkham}. 
The second derivative ($d^{2}R/dB^{2}$) was calculated in order to remove the regular field-dependent magnetoresistance from the raw data, 
leaving out only the L-P oscillations. Interestingly, the A-B effect is observed only when the magnetic field $B$ is swept from the largest (1 T) 
to its lowest value (-1 T). The period for the L-P oscillations is $B_{L-P} \simeq 0.2$ T. 
This period is close the $\simeq 0.15$ T period found for the oscillatory magnetoresistance at a
topological insulator/chalcogenide interface \cite{Dean}.   
\begin{figure}
\centering
\includegraphics[width=3.2in]{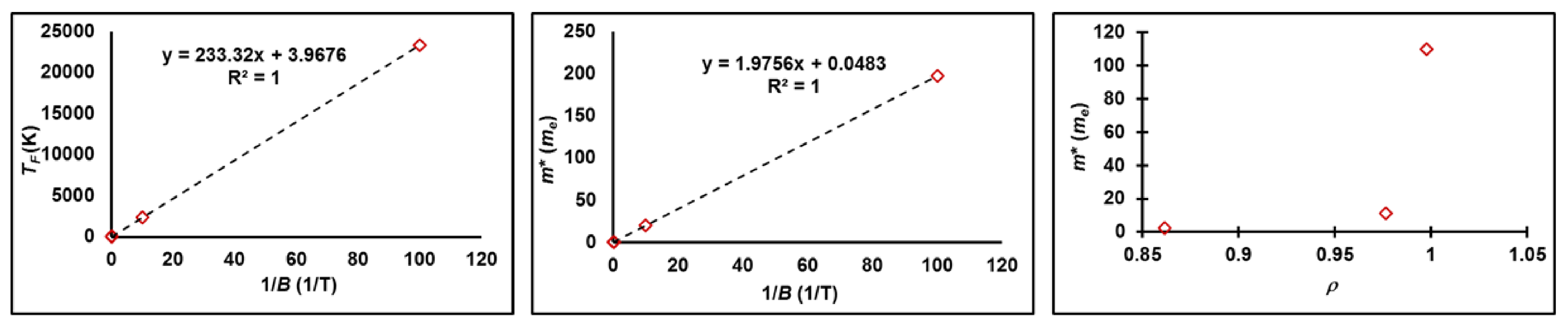}
\caption{Fermi temperature $T_{F}$ and the effective mass $m^{*}$ vs. inverse field $1/B$. 
The effective mass $m^{*}$ vs. the band filling factor $\rho$.}
\label{Fig15}
\end{figure}
The L-P oscillations, which reflect the pattern for the amplitude and phase of the order parameter, 
were found to be essentially independent of sample dimensions, temperature, transport
current, and the magnitude and orientation of the magnetic field, including magnetic fields oriented
parallel to the film plane \cite{Kunchur}. This situation would suit well the C fiber, where the many graphitic planes have equally 
many orientations with respect to a field normal to the fiber's axis.
We also notice that $\mu_{B-P}B_{L-P} \simeq 11.6$ $\mu$eV in this case ($\mu_{B-P} \simeq 0.927 \times 10^{-23}$ J/T is 
the Bohr-Procopiu magneton) would be about two times the Lamb shift for H atom for which there is the split
between the $2s_{1/2}$ and $2p_{1/2}$ energy levels is $\simeq4.38$ $\mu$eV.
The Lamb shift is related to the coexistence of SC and magnetism, including the coexistence with an incommensurate
spin-density wave (magnon) at a structural transition to an anisotropic phase (in the $ab$ plane).
This $11.6$ $\mu$eV value is also closed to the smallest gap $\Delta$(130 K) $\simeq 10$ $\mu$eV (Fig. \ref{Fig9}).
Does this mean that the spin wave survives past $T = 130$ K to at least $T = 300$ K?
More importantly, the L-P oscillations were observed with the magnetic field directed along the SC cylinder’s axis.
How that we are observing L-P oscillations with the magnetic field normal to the axis of the octane-intercalated C fiber?
Here is what we believe that is going on. First, let us apply the same reasoning as the one used for the original L-P oscillations 
to find for ``corresponding cylinder/fluxon radius'' the octane-intercalated C fiber:
$R = \sqrt{\frac{\Phi_{0}}{\pi B_{L-P}}} \simeq 56\hspace{0.1in}\textnormal{nm}$.
where $\Phi_{0}=h/2e \simeq 2.1 \times 10^{-15}$ T $\cdot$ m$^{2}$ = $A \times B_{L-P}$ is the magnetic flux quanta. 
From here, the fluxon area in this case is $A \simeq 1.0 \times 10^{-14}$ m$^{2}$.
Rather than only one fluxon, a more favorable energetic situation is to have $N$ fluxons, each having a radius $R/\sqrt{N}$ 
and magnetic flux quanta $\Phi_{0}/N$ \cite{Tinkham}. In a microring of a SC with a spin-triplet equal-spin pairing state, 
a fluxoid, a combined object of magnetic flux and circulating supercurrent, can
penetrate as half-integer multiples of the flux quantum \cite{Sharon}.
Chirality-controlled spontaneous currents in spin-orbit coupled superconducting rings have been observed \cite{Robinson}.
Instead of an integer number of fluxons, is it possible that the octane-intercalated C fiber is a chiral $p$-wave SC \cite{Volovik1} 
harboring a $N = 1/8$ fractional fluxon inside a M\"{o}bius SC strip? Then the corresponding area would need to be 
$A \simeq 1.3 \times 10^{-15}$ m$^{2}$ giving for the fluxon's radius $R_{1/8} \simeq 20$ nm. 
Here is a possible scenario for the octane-intercalated C fiber: In the absence of the field, the system is described by the Little model for
high-$T_{c}$ organic SC. The spines are in the $ab$ plane (the graphite plane).
When a magnetic field (in this case $B = 0.2$ T) is applied normal to the C fiber’s axis, the SC condensate is containing a 1/8 fractional fluxon.
Others observed L-P oscillations in a chiral SC nanotube \cite{Qin1}.
It is possible that the octane-intercalated C fiber contains chiral noncentrosymmetric tubular nanostructures.
The noncentrosymmetric property is a result of mixing the of the singlet and triplet states (i.e., different
parity states) \cite{Gorkov}. It was also found that for WS$_{2}$ nanotubes, $T_{c}$ increases with the diameter of the tube \cite{Qin2}.
The LP effect originate from the interference of the SC current along the nanotube circumference and the resultant oscillations of $T_{c}$.
Given all these, is there a more likely to consider that the magnetic flux area is planar and when it extends over the fiber's length
$l = 1.4$ mm, then it must be $w = A/l \simeq 7.7 \times 10^{-12}$ m wide. 
This is quite a very small dimension, the like of thread bonds in molecules \cite{Ivlev}.
If the model proposed by Little \cite{Little1} is used with the polarized octane molecules alternating on the sides of the SC spine, 
then the radius of the thread bond is the distance between the two H atoms at the end of two octane molecules on opposite sides of the spine. 
The width of the SC spine is $w_{SC spine} = 2 \times l_{C−H} = 2 \times 1.1$ \AA $ = 2.2$ \AA.
Because the distance along relatively straight portions of the PAN-fibers’ graphitic network is on average not larger than 
$l \simeq 10$ nm, the more likely situation is that the magnetic flux area is $A = d \times w$, where $d = 7.0$ $\mu$m is the fiber's diameter.
For one fluxon, the width for the flux area is $w = 10^{-14}$m$^{2}/d \simeq 1.4$ nm. That is, the coherence length at $T = 300$ K 
is $\xi_{m} \simeq 1.4$ nm.

Using the $T_{c} \simeq 440$ K value estimated from the BCS fit of the gap data and the temperature dependence 
$\xi(T) = \xi_{0}\sqrt{1-(T/T_{c})^{2}}$, we find for the zero temperature coherence length $\xi_{0} \simeq 1.9$ $\mu$m.
Then the electron density parameter is $r_{s}=\xi_{0}/2a_{B} \simeq 18$, with $a_{B}=0.529 \times 10^{-10}$ m the Bohr (H atom) radius.
Interestingly, the product $r_{s}m^{*} \simeq 40$ is the same for BSCCO, where $r_{s} \simeq 2.5$ and $m^{*} \simeq 16$ \cite{Bianconi}.
What could be the reason for the product $r_{s}m^{*}$ to be constant? From $\xi_{0}=2r_{s}a_{B}$ and $\xi_{0}=\hbar v_{F}/\pi \Delta(0)$,
we find $\Delta(0)\frac{2\pi a_{B}}{v_{F}} \le \hbar$ or $\Delta(0)\frac{l_{B}}{v_{F}} \le \hbar$, where $l_{B}=2\pi a_{B}$ is the Bohr
circumference in the Wigner localization limit. Not surprisingly, we have recovered the Heisenberg relation with $\hbar = h/2\pi$ 
the reduced Planck constant. The time lapse for completing the $l_{B}$ track at Fermi velocity $v_{F}$ or the Bohr period is $T = l_{B}/v_{F}$ , 
from where the angular velocity is $\omega_{B}=2\pi v_{F}/l_{B}$. Using the values at $T = 300$ K
and $B = 0.2$ T, $v_{F} \simeq 8.0 \times 10^{4}$ m/s and $l_{B} = 2\pi \times 0.529 \times 10^{−10}$ m, we find
$\omega_{B} \simeq 1.5 \times 10^{15}$ Hz, which is close to the angular frequency for the
1.6 eV exciton, $\omega_{exciton} \simeq 2.4 \times 10^{15}$ Hz.
This result suggests the existence of a SC condensate formed by the partial density of the
Fermi liquid close to the shape resonance condition that is coupled with the 1D generalized Wigner polaronic charge density wave
(GWP-CDW) like the one found in the BSCCO HTS compound \cite{Bianconi}. These results can be also correlated to the behavior of (the
resistivity) $\rho$ at $T \simeq 250$ K, where either a I-M transition or a M-I is observed depending on the excitation current 
20 $\mu$A or 1 $\mu$A, respectively.

The evidence of L-P oscillations in the octane-intercalated C fiber suggests the existence of a macroscopic quantum SC phase at
temperatures at least 300 K. To our knowledge, this is the second time time that the L-P oscillations are reported in a graphitic material \cite{Gheorghiu}.
As the macroscopic quantum phase results from the condensation of quasiparticles, we have also estimated the quasiparticles's
effective mass $m^{*} = \sqrt{2E_{F}/v_{F}^{2}}$, where the Fermi energy $E_{F} = k_{B}T_{F}$ was found from fitting 
the $\sigma(T)$ data (Tab. \ref{m*}) to Eq. \ref{Koike_eq}. The band filling factor $\rho$ \cite{Kim} was also found.
As Fig. \ref{Fig15} shows, the dramatic increase of $m^{*} > 10^{2}m_{e}$ suggests that the octane-intercalated C fiber is a heavy fermion HTS.
Since the plasmon's energy is  $\Delta_{plasmon} = -W_{Casimir}$, thus $\omega_{exciton} \simeq 9.5 \times 10^{14}$ Hz is much larger than the 
plasma frequency $\omega_{plasmon} \simeq 6.2 \times 10^{12}$ Hz. This is what might have happened: An exciton is formed at $T \simeq 50$ K. 
As the temperature is increased to room-$T$, the extend of the exciton (electron-hole pairs) reduces and exciton breaks into the more dynamic bipolarons.
For $B = 0.2$ T, the magnetic length is $r_{B} \simeq 1$ mm, i.e., $r_{B} = \sqrt{\hbar c/eB}$ scales with the 1.4 mm length of the C fiber. 
Notice also that for $B = 0.2$ T, $m^{*} \sim 2m_{e}$. It was found that $m^{*} \simeq 2m_{e}$ is the optimal value
that can give a high $T_{c}$ in all cuprates when the Casimir effect is considered \cite{Orlando}.
The Casimir effect was recently found relevant also for graphene \cite{Bortag}.
Can the Casimir effect be relevant in our case? The Casimir energy is given by $W_{Casimir} = -\frac{\hbar c\pi^{2}A}{720d^{3}}$,
where $c=\simeq 3.0 \times 10^{8}$ m/s is the the velocity of light in vacuum, $A$ is the area of each of the two Casimir planes and
$d$ is the small distance between the planes. 
\begin{table}
\centering
\textsc
\bigskip
  \begin{tabular}{|c|c|c|c|}
\hline
  $B$(T) & $T_{F}$(K) & $m^{*}$($m_{e}$) & $\rho$ \\
\hline
  0.01 & 23200 & 110 & 0.998 \\
  0.1 & 2330 & 11.0 & 0.977 \\
  0.2 & 471 & 2.23 & 0.862 \\
  5  & 49.9 & 0.236 & 1.34 \\
  9 & 27.8 & 0.132 & 1.60\\
\hline
  \end{tabular}
\caption{Fermi temperature $T_{F}$, effective mass $m^{*}$ (in units of the electronic mass $m_{e}$), 
and the band filling factor $\rho$ in different fields $B$.}
\label{m*}
\end{table}
We can also estimate the 2D carrier density for each parallel 2D conduction channel, $n_{2D} = g|e|B_{L-P}/h \simeq \sim 10^{-14}$ m$^{-2}$, 
where $g = 2$ is the spin degeneracy,
$e = 1.6 \times 10^{-19}$ C is the electron charge, and $h = 6.62 \times 10^{-3}$ J$\cdot$s is the Planck constant.
The 3D carrier density is given by $n_{3D} = \sigma/(|e|\mu)$, where $\sigma =1/\rho$ is the electrical conductivity
and $\mu = v_{F}/E = v_{F}/(V/d)$ is the electron mobility, $v_{F}$ is the electron velocity at the Fermi 
level, $E$ is the strength of the electric field, $V$ is the potential difference, and $d$ mm is length of the fiber.
Using the experimental data for $\sigma = 5.9 \times 10^{4}$ $\Omega^{-1}\cdot$m$^{-1}$, $V = 2.5 \times 10^{-3}$ V, and
$d = 1.4$ mm, the Fermi velocity $v_{F} = \sqrt{E_{F}/2m^{*}}$ with the Fermi energy $E_{F} = k_{B}T_{F}$ with $T_{F} \simeq 397$ K
found from  fitting the $\sigma(T)$ data to the Koike formula 
$v_{F} \simeq 5.3 \times 10^{4}$ m/s, we find $\mu \simeq 3.0 \times 10^{4}$ m$^{2}\cdot$s$^{-1}\cdot$V$^{-1}$
and $n_{3D} \simeq 1.2 \times 10^{19}$ m$^{-3}$. Then the effective thickness per 2D conduction channel, given by the
ratio $n_{2D}/n_{3D} \simeq 8.3$ $\mu$m, is very close to the fiber's diameter $\simeq7$ $\mu$m.
That is, \textit{the octane-intercalated C fiber is a topological object}.

\section{Conclusions}
We have continued our work on finding SC in C allotropes \cite{Pierce},
herein considering C fibers modified by intercalation with octane.
The reentrant I-M transitions are more than just anomalies in the $\rho(T)$.
Below $T \sim 50$ K, the magnetoresistance data shows a transition from AFM to FM correlations as the
strength of the magnetic field is increased. 
Low-temperature PM is also observed. Just as in cuprates \cite{Stepanov}, it is possible that in these hydrogenated graphitic system
collective spin fluctuations are well developed even in the PM regime. Superconductivity might be mediated by 
both short-range AFM correlations and long-range FM correlations, where the latter become dominant as $B$ is increased. 
Nonoverlapping AFM and FM and overlapping AFM and SC as well as FM and SC have been discussed in \cite{Zaitsev}.
The irreversibility observed in the FC vs. the ZFC data for a sufficiently high magnetic field suggests that 
the system enters the SC state below $T_{c} \sim 50$ K. 
Thus, the octane-intercalated C fiber appears to be an unconventional magnetic SC.
In addition, AFM spin fluctuations creates pseudogap states above $T_{c} \sim 50$ K \cite{Schrieffer4}.
It is also possible that $T \sim 50$ K is a tricritical point, with the magnetic field acting as a chemical potential that can turn the low-$T$ AFM into FM. 

The unlikely coexistence of SC and FM was observed in bulk insulators like SrTiO$_{3}$ \cite{Koonce,Dikin}. 
Recently, a model was proposed in order to explain the magnetic-field driven SC-I quantum phase transition in disordered Josephson chains \cite{Bard}. 
The SC-I transition arises naturally in low-dimensional systems (1D, 2D), as the Anderson localization
precludes, in general, the emergence of an intermediate metallic phase. Large DOS at the Fermi level owning to the presence of
edge states and topological disorder can trigger competition between SC instabilities and charges density waves in these quasi-2D C materials. 
\textit{Topology} is the key word in understanding the I-M transitions \cite{Khodel} and the SC-like 
behavior \cite{Khodel,Esquinazi3} observed in these C fibers.
The octane's work on the graphitic planes might favor 2D SC, with H playing an important role.
In metallic systems, the grains must be in close contact for the field-induced percolation to be effective \cite{Abeles1}.
Electron microscope studies show that the microstructure consists of a wide distribution of grain sizes, with some of the grains surrounded by 
insulator and chains of touching grains. While bulk SC and FM are vanishing abruptly 
below the percolation threshold, an individual grain can retain SC if its size is larger than the coherence length $\xi_{c}$ \cite{Abeles2}.
Likewise, the many new interfaces created by octane within the turbostratic volume of the C fiber can host SC \cite{Esquinazi2}.
Magnetization measurements on heptane-treated high-quality
graphite powders suggest the existence of granular SC at certain interfaces, which are possibly
created after the treatment of graphite with the alkane \cite{Esquinazi2}. One question is whether the alkane is
creating 2D grain boundaries that are normal to the graphite planes and perhaps the H ions (H$^{+}$ or protons) in the alkane 
are responsible for the observed phenomena.
For instance, experiments found that the treatment with octane can lead to free protonation of the graphite's interfaces \cite{Kawashima1} 
and the observation of HTS in C fibers \cite{Kawashima2}.

We have further corroborated the magnetoresistance results with the energy gap data as obtained from nonlocal $G_{diff}$ measurements. 
An excitonic mechanism is likely driving the system to a SC state below the same $T \sim 50$ K, where the gap is divergent.
The temperature-dependent gap $\Delta(T)$ as obtained from nonlocal $G_{diff}(V)$ experimental data 
for the octane-intercalated C fiber suggest possible transitions to SC, including HTS states occurring 
in a multiple-gap system with critical temperatures estimates above room temperature. 
The HTS phases are likely surface SC states formed locally at the grain boundaries of the octane-intercalated fiber,
similarly to what has been found at the Cu-CuO interface \cite{Osipov}.
Thus, the magnetoresistance and the conductance measurements here show that
either a magnetic or an electric field can drive this hydrogenated graphitic system to a SC state below $T_{c} \sim 50$ K. 
The temperature dependence of the SC gap follows the flat-band energy relationship.
We have also found that the flat band gap parameter has a linear increase with with the temperature above $T_{c} \sim 50$ K.
The Delta-function-like peak observed at $T \sim 50$ K in the energy gap $\Delta(T)$ for the octane-intercalated C fiber 
is a sign of soliton formation. It was suggested that topological HTS, which can be found also in organic compounds, 
occurs due to the formation of solitons \cite{Schrieffer2,Schrieffer3}.
Unlike the electrons from conventional SC, the Cooper electrons in cuprates and some other unconventional HTS are pairs
of soliton-like excitations. Solitons play a fundamental role in the charge-transfer doping mechanism.
Bisolitons, i.e. electron-hole excitations or polaronic solitons, might lead to the HTS mechanism in ceramic oxide materials like YBCO 
with the SC generated by quasi-1D chains (molecular-like) of alternating Cu$^{2+}$ and O$^{2-}$ ions \cite{Davydov1,Davydov2}.
The polaronic (or bipolaronic) mechanism is fundamental to the formation of Cooper pairs in HTS materials \cite{Andreev,Muller}.
The bisolitons, which are Bose-particles with zero spin and a double electric charge, form a Bose-condensate usually at low $T$. 
The moderately strong, nonlinear electron-phonon interaction is responsible for electron pairing in the cuprates, 
where the perturbation theory becomes invalid.
Thermal vibrations of the lattice atoms lead to an increase in $2\Delta(0)$/$2k_{B}T_{c}$ to values that might exceed the BCS one ($\sim 3.53$), 
in accordance with experiments. At the same time, consider the case of a long-range phase coherence, like the magnon-induced SC in a
topological insulator coupled to FM and AFM insulators \cite{Hugdal}. The pairing strength also increases with increasing 
Fermi momentum, in agreement with previous studies in the limit of high chemical potential $\mu$. 
The pairing strength decreases as $\mu$ moves towards the gap and completely vanishes when $\mu$ is inside the gap. 
If a quadratic term is present in the Lagrangian, the pairing is instead BCS type up to a certain value of $\mu$. In the AFM case, the BCS pairing
occurs when the FM coupling between magnons on the same sublattice exceeds the AFM coupling between magnons on different sublattices. 
Outside this region in parameter space is the Amperean pairing leading to SC.

Perhaps relevant for this work is the prediction that $p$-doped hydrogenated graphene (or graphane) is an electron-phonon SC with a $T_{c}$ larger than 90 K \cite{Savini}. 
The main difference between graphene and graphane is that, while the former is fully sp$^{2}$ bonded, the latter is sp$^{3}$ bonded, as diamond is \cite{Sofo}.
Since graphane is the 2D counterpart of diamond, which is SC when $p$-doped, calculations show that doped graphane is a potential HTS material \cite{Savini}.
Graphane is also one of the most stable hydrocarbons. It is possible that the octane intercalation results in the formation of graphane domains within the sample, 
thus making possible HTS. 
The DC magnetization measurements that we have conducted on other octane-intercalated graphitic samples, such as graphite powder 
and the HOPG-Kapton tape-HOPG composite, have confirmed the existence of SC properties in these hydrogenated graphites.

One important question is the nature of the SC-like behavior in these C fibers.
For a BCS-type SC, the $T_{c}$ is given by Eq. \ref{McMillan}:
With the $T_{D} = 2430$ K, room-$T$ BCS-like SC in graphite is achieved for $N(0)\lambda \approx 0.45$.
A BCS SC like Pb, for which $T_{D} = 96$ K, has $N(0)\lambda \approx 0.39$ and $T_{c} = 7.2$ K. 
The relation between the Fermi temperature $T_{F} = E_{F}/k_{B}$ and the Debye value $T_{D}$ plays a more subtle role.
In the theoretical approach for HTS materials, $E_{F}$ is considered a variable. On the contrary, $E_{F}$ is not 
a variable in the BCS theory due to the assumption that $E_{F} \gg k_{B}T_{D}$, or $T_{F} \gg T_{D}$ \cite{Malik}.
While not included with this work, estimates of the Fermi energy $E_{F}$ for the octane-intercalated C fibers will be the topic for a follow-up paper.
The HTS mechanism in unconventional materials such as the cuprates and the C-based cannot be entirely phonon-based,
as it needs long-range electronic correlations that are not phonon-mediated. 
In cuprates, long-range phase coherence results from the interplay between the lattice and spin fluctuations.
In C-based SC, as the octane-intercalated C fibers here, SC might arise via the excitonic mechanism \cite{Little1,Ginzburg1,Allender}.
The indirect exciton at 50 K is the driving force for the boson condensation and the BKT transition below 50 K.
High-temperature superfluidity with indirect excitons in van der Waals heterostructures have been studied before \cite{Fogler}.
As shown by this work, the observed SC-like features of alkane-intercalated C fibers hint to an underlying excitonic mechanism that 
can be explained within the framework of FB energy bands. Inside the spin-triplet excitonic phase and at finite chemical doping, a pairing interaction, 
conducive for SC order, may arise from AFM fluctuations \cite{Roy}. 
Ginzburg considered the possibility for the pairing of electrons in metal layers sandwiched between polarizable dielectrics through virtual excitations at high energy.
In graphene, excitonic resonances are a signature of Coulomb (Amperean) coupling \cite{Stroucken}, seemingly the SC
mechanism in FMSC, i.e., spin-triplet SC \cite{Lee1}. Contrary to the Cooper pairing, the Amperean pairing is an
\textit{attractive} interaction between fermions with parallel momenta that occurs whenever the effective Lagrangean for the magnetic fluctuations does not
contain a quadratic term \cite{Kargarian}. The Amperean mechanism for SC has been already proved in the case of magnon-induced SC in a topological insulator 
coupled to FM and AFM insulators \cite{Hugdal}.
The FM and the excitonic signatures observed with the octane-intercalated C fibers are a hint for possible Amperean coupling in this system.

We believe that our results and their interpretation are contributing to the mounting evidence of unconventional
SC located at graphite's interfaces, in particular after the samples were brought in contact with alkanes.
This work is among the few reporting on HTS-like states in hydrogenated graphites.
We are also inquiring on the possible coexistence of surface SC and magnetism \cite{Ginzburg2} at interfaces of graphite that had undergone
certain modification processes. It is suggested that in order to reach room-$T$ SC, one must search for or artificially create systems 
that experience the nontopological FB in the bulk or topologically protected FB (zero-energy bands) on the surface or at the
interfaces of the samples \cite{Heikkila}. The larger than in the bulk $T_{c}$ is due to its proportionality to both
the coupling constant and the area of the FB.
Recently, unconventional SC below $T_{c} = 1.7$ K was found in 2D superlattices obtained by stacking 
two graphene sheets at a 'magic' twist angle 1.1$^{0}$ \cite{Cao}. 
Simple calculation finds that approximately $\sim 330$ graphene layers would be needed to accomplish a complete $360^{0}$ twist,
amounting to $\sim 57$ nm thick sample. Perhaps more than just coincidentally, in bulk HOPG samples a semiconducting-metallic transition 
occurs at a minimum sample thickness of $\sim 50$ nm \cite{Garcia1}.
Another more than just a mere coincidence is that the magic angle $1.1^{0}$ by which the graphene layers need to be twisted with respect 
to each other for SC to arise is close to the one related to a problem that has fascinated researchers from Kepler, to D'Arcy Thompson, and Hales \cite{Adam}.
Kepler asked: What is the most efficient partition of space into equal volumes? 
Hales reformulated the problem as: Any partition of the plane into regions of equal area has the perimeter at least 
that of the regular hexagonal tiling \cite{Hales}.
I.e., the honeycomb tiling is the most packed tiling of all. D'Arcy Thompson gave a twist to the problem:
Imagine a hexagonal prism, open at one end and sealed at the other by three rhombuses. What is the apex angle that minimizes the total surface area of the cell?
It turns out that that angle is about $1^{0}$. What all this diversion tells us is that 
the understanding of the packing of the honeycomb structure is no less important
than the honeycomb symmetry itself for the class of SC materials and heterostructures. If nothing else, it also tells us that the bees are right.  
In bulk systems, the graphitic domains should be twisted with respect to each other in order to form the FB in electronic spectrum. 
The topological origin of the FB can be also understood in terms of the pseudo-magnetic field created by strain \cite{Volovik2}. 
On a more fundamental level, finding SC in C-based materials should not come as a surprise. 
London was first to suggest quantum behavior in the form of superfluidity in organic matter \cite{London}.
Compounds with aromatic rings such as anthracene and naphtalene might exhibit a current running freely around
the rings under a charging magnetic field. 
The formation of a Cooper pair approximately 40 eV above the double-ionization threshold in benzene, naphthalene, anthracene, and coronene 
after the absorption of a single photon was also observed \cite{Wehlitz}.
Little \cite{Little1,Little2} was the first to suggest the possibility for room-$T$ organic SC with a  maximum $T_{c} \sim 2200$ K.
Charge carriers moving along a conducting spine are bound via the virtual electronic excitation of polarizable side groups.
Interestingly, both living forms of organic matter (mostly conjugated hydrocarbons, i.e., C and H atoms) and oxide HTS have unpaired spins and own their 
function on the presence of O$_{2}$. In addition, while the $\sigma$ electrons are localized close to the nuclei and thus 
do not differ much from the ordinary atomic electrons, the $\pi$ electrons move collectively in the field created by the former.
Hence, the $\pi$ electrons' behavior as conduction electrons is making their host, the hydrocarbon molecules, 
similar to metals. Moreover, it turns out that the hydrocarbons with even number of C atoms are small SC \cite{Kresin1}.
The pair correlation, which takes place in a number of Fermi systems, can lead to SC in metals and 
in $\pi$ electronic systems also leads to Josephson-like phase coherence. 
Thus, the occurrence of both pair correlation and the field-induced metallic states in the octane-intercalated C fibers
suggests similarity between the system here under investigation and the family of metallic nanoclusters, with the latter considered
potentially HTS materials \cite{Kresin2}. The pairing in the nanoclusters is similar to that in the atomic nuclei \cite{Pines}-\cite{Zelevinsky}. 

We have found evidence of superconductivity in hydrogenated graphenes.
Recent studies show the viability of life in H$_{2}$-dominated exoplanet atmospheres \cite{Seager}.
Thus, \textit{it appears that both superconductivity and life thrive on hydrogen}. \textit{Maybe that life needs superconductivity?}

\vspace{0.1in}
\centerline{\textbf{ACKNOWLEDGMENTS}}
\normalsize
\vspace{0.05in}
This work was supported by The Air Force Office of Scientific Research
(AFOSR) for the LRIR \#14RQ08COR \& LRIR \#18RQCOR100 and the Aerospace Systems Directorate (AFRL/RQ). 
We acknowledge J.P. Murphy for the cryogenics. 
First author's special thanks go to Dr. G.Y. Panasyuk for his continuous support and inspiration.\\

\end{document}